\documentclass[11pt,aps,prd,eqsecnum,superscriptaddress,a4paper,longbibliography]{revtex4-2}
\usepackage{amssymb,amsmath,amsthm,graphicx,amscd}
\usepackage{enumerate,comment,ulem,bm,enumitem}
\usepackage[mathscr]{eucal}
\usepackage[cal=boondoxo]{mathalpha}
\usepackage{bm,pbox,cancel}
\usepackage{wrapfig,subcaption}

\usepackage{xcolor}
\usepackage{etoolbox}

\renewcommand{\bibinfo}[2]{%
    \ifstrequal{#1}{title}{\textit{#2}}{%
        \ifstrequal{#1}{booktitle}{\textsl{#2}}{#2}%
    }%
}

\usepackage[hidelinks]{hyperref}
\usepackage{orcidlink}

\makeatletter
\renewcommand*\env@matrix[1][\arraystretch]{%
  \edef\arraystretch{#1}%
  \hskip -\arraycolsep
  \let\@ifnextchar\new@ifnextchar
  \array{*\c@MaxMatrixCols c}}
\makeatother

\DeclareMathSymbol{\shortminus}{\mathbin}{AMSa}{"39}

\begin{document}

\title{Motional Decoherence of the internal state of an atom by vacuum and thermal fluctuations of a quantum field: Necessity of a fully relativistic treatment}
\author{Jen-Tsung Hsiang\orcidlink{0000-0002-9801-208X}}
\email{cosmology@gmail.com}
\affiliation{College of Electrical Engineering and Computer Science, National Taiwan University of Science and Technology, Taipei City, Taiwan 106, R.O.C.}
\author{Bei-Lok Hu\orcidlink{0000-0003-2489-9914}}
\email{blhu@umd.edu}
\affiliation{Maryland Center for Fundamental Physics and Joint Quantum Institute,  University of Maryland, College Park, Maryland 20742, USA}
\date{November 2, 2026 version}

\begin{abstract}
We consider a massive object, such as a neutral atom or an Unruh-DeWitt detector, moving at constant speed in a quantum field and investigate how motion affects the quantum coherence of its internal degrees of freedom, manifesting as quantum dispersion in the absence of a quantum field, and decoherence in its presence. Our investigation aims to serve two purposes: One is to obtain the correct results for these effects, which are essential for the design of quantum information processing schemes where the quantum coherence of an atom's internal degrees of freedom is of major concern. The other is to expose the fallacies in commonplace approaches, which start with a non-relativistic calculation and then add on relativistic ``corrections". We first show a fully relativistic calculation of the internal dynamics based on covariance principles, and then one which ignores these principles, often justified by the atom's slow motion.  By a direct comparison of results from these two approaches, we identify  {significant discrepancies} in a non-relativistic treatment, even at  low atomic speeds. The proper procedure, and a much safer one in terms of accuracy, is recommended here:  first derive the dynamics and the necessary regularizations within a fully covariant relativistic framework. Only after the exact, internally consistent equations of motion are established should low-velocity limits be extracted from them.  Only in this way can all the relativistic effects which matter: time dilation, aberration, non-inertial mass shifts, be systematically preserved, and accurate results obtained. 
\end{abstract}
\maketitle

\newpage
\tableofcontents

\section{Introduction}

In this paper, we consider a massive object with internal degrees of freedom (idf) carrying a charge, such as a neutral atom with electronic activities, or an Unruh-DeWitt (UdW) detector \cite{Unr76,DeW79} with mass \cite{SudKem,Zych}, an idealized point-like object with idf, moving at constant speed, interacting with a free (non-interacting) quantum field in a space with no boundary.  We aim at finding out a) how the quantum field influences the atom’s idf,  manifesting as \textit{quantum dispersion}, even when it is at rest, and as \textit{motional decoherence}, when it is in motion. We use the variance of the displacement of the idf, similar to the spread of a wavepacket, as a measure. We consider both quantum and thermal effects by treating the ambient quantum field at zero or at finite temperatures.   

\subsection{Motivation, Issues and Goals}

What motivated us to take up this study is two-fold:  From a foundational theoretical perspective, this is perhaps the most basic level of inquiry into the quantum nature of atom-field interactions. From the simplest situation of a  quantum object sitting still,  represented by a wavepacket in quantum mechanics textbooks, whose wavefunction spreads in time without any interaction and in the absence of an environment, known as \textit{quantum dispersion}, to that of a moving quantum object with idf which interacts with an ambient quantum field, leading to an increase in the variance of the idf wave function, known as \textit{motional decoherence}. At a more practical level of inquiry, motional decoherence affects the activities of the atom's internal states, such as spontaneous emission or Rabi oscillations   \cite{Shresta,Bernad,RabiOsc}, which in turn can influence the kinematical state of the external or mechanical degree of freedom (mdf), such as quantum friction \cite{Viotti1,Viotti2}. Note the difference between motional decoherence of the internal dof studied here and the spatial or path decoherence studied by many authors \cite{MazzitelliPaz2003,HsiangFord2004,HsiangLee06,Knight,Scheel,Nori,Henkel,Kanu}. The latter addresses the external dof: the change in the spread of the quantum wave function of an atom towards its classical value in displacement. Because many modern techniques in controlling atoms are based on manipulating the interaction of the atom's idf with a quantum field, often a laser designed to perform specific functions, motional decoherence is an important factor of concern for high accuracy  targeting and positioning of neutral atoms in traps or optical lattices \cite{Nobel97} and for ion-trap  quantum information  processing (QIP) designs. The interplay between the motional and internal states of an atom was analyzed in full seriousness for the necessary feasibility studies very early on \cite{meekhof1996,Wineland,Blatt}.  

For our purposes, this is also a necessary step towards uncovering the basic mechanisms for possible \textit{vacuum viscosity} effects such that a massive object may experience when moving in an ambient quantum field. This is related to the familiar topic of \textit{quantum friction}, the root cause of which is still unclear to us and is what we aim to explore in a follow-up study. The focus there would be on the mechanical or external degree of freedom (mdf), i.e., the center of mass of the atom or the UdW detector.  We ask whether the backreaction of the vacuum fluctuations in the quantum field through its interaction with the idf would engender non-Markovian dissipative effects (with memory) on its mdf. We have carried out such a study for classical fields only \cite{HHVacVisCF}, drawing attention to the necessity of a fully relativistic treatment. Full treatment is in contrast to hybrid treatment, meaning, not only for the quantum field, which is well known, but also using proper time, not the laboratory time, for the idf dynamics. Describing the idf dynamics with nonrelativistic quantum mechanics is almost standard in atomic-molecular-optical (AMO) physics, but the violation of Lorentz invariance in any kind of hybrid treatments could lead to nonphysical effects such as the appearance of fictitious friction \cite{Wilkens,Barnett,Sonnleitner,Wood}.   The focus in our treatment of vacuum viscosity in a sequel paper would be on quantum noise from the fluctuations of quantum fields at zero temperature, as the thermal field effect would manifest in the much better understood Einstein-Hopf drag \cite{EinHop}.   

\subsection{Modeling based on microphysics dynamics}

We adopt a microphysics model for quantum optomechanics, the so-called AMOF model \cite{Galley13,Sinha15,Lin18,SLS21,Butera23,MOF6,AFM1,AFM2},  to describe the dynamics of a massive object with internal structure. It comprises three interlinked dynamical variables: i)  its center-of-mass $M$ (in a nonrelativistic setting), which we refer to as the mechanical (or external) degree of freedom (mdf),  moving along a spacetime trajectory $z^{\mu}$ and ii) an internal degree of freedom (idf)  $\chi$, modeled here as a harmonic oscillator with mass $m$ and  charge $\lambda$, interacting with iii) an ambient massless scalar field $\phi$. 

Since the object is treated as a pointlike entity, $\chi$ and $\phi$ are bilinearly coupled  along the object's worldline, meaning, the local field dynamics that drives the internal dynamics are evaluated  at $z^{\mu}$. This framework is an effective model for a neutral atom: we can call it a two-level atom if two-level activities are the focus of attention, or a harmonic atom if the internal space variable $\chi$ undergoes harmonic motion,  representing an induced, non-permanent dipole. 

\paragraph{Simplifications for highlighting the essentials}

For simplicity, we work with a scalar field here coupled with a monopole charge representing the internal dof. In the real world situation, what this model emulates would be for electrons in an oscillating dipole distribution interacting with an external electromagnetic field in a derivative type of coupling.  The difference from a full treatment of a real neutral atom is our ignoring the  positive charge carried by the nucleus in balance with that of the electron. Since our focus here is on how the idf interacts with an external field, and in a sequel study, how that affects the motion of the mechanical dof, the mass $M$, we shall leave out the interaction between the charged mass $M$ and the field, which has  little effect on the issues at hand. 

A relativistic description is needed for the treatment of the changes to spatial and momentum distributions, spin–orbit interactions, quantum electrodynamic corrections such as the Lamb shift, and vacuum polarization. Our present model picks out only the interactions of the idf with an ambient quantum field assumed to be the total field in its vicinity. The idf in our modeling is not an electron with spin but a harmonic oscillator, and the field is not an EM field with polarizations, but a  quantum scalar field at zero and finite temperatures, since the important physics we want to illuminate on originates from its vacuum and thermal fluctuations.  This simplified model serves the function of what we aim at here, much as the highly idealized Unruh-DeWitt detector model with a two level or a harmonic oscillator idf focusing on its interaction with a quantum scalar field brings out the famous Unruh effect when the atom is in uniform acceleration.   

\subsection{Deeper issues not addressed here}

Relativistic effects for a real atom is much more involved, and approximations are often invoked from the full QED theory, which treats electrons as Dirac fermions and the binding of the electrons to the nucleus as a bound state described by Bethe-Salpeter equations, etc.  One level down is relativistic atomic structure theory,  which  treats the atom or molecule as an assembly of electrons and atomic nuclei interacting primarily by exchanging photons. Even with this model, simplifications are required, such as representing the nuclei as classical charge distributions, or even as point particles. (This is obviously inadequate for heavy atoms, where the electric field gradient requires relativistic calculations of the nuclear quadrupole moments \cite{Ilias}.) Since the motion of the nuclei is usually slow relative to the electrons, it is often adequate to treat the nuclear motion with respect to the electrons nonrelativistically, or even to start from nuclei in fixed positions, correcting subsequently for nuclear motion \cite{Grant}.  However, for atoms or ions moving at relativistic speed, even the Lorentz contraction of a bound state as simple as a hydrogen atom, is not a straightforward calculation \cite{jarvinen}. In this study, we assume that the position of the atom is the same as that of its nucleus,  regarded as a point particle with mass $M$, which we refer to as the atom's external or mechanical dof (mdf), moving at a prescribed trajectory, and in this paper, with uniform relativistic speed. The word `prescribed' means it is given and stays as given,  in contrast to the case where the trajectory is determined by including the backreaction of the quantum field on the mdf through its action on the idf, as considered in \cite{HHVacVisCF,HHVacVisQF}. The word ``fully relativistic" in the title should not be taken to mean `complete' but refers specifically to treating both the quantum field relativistically as it should be, and also the idf dynamics, the internal state, in terms of proper time, in both the laboratory and the comoving frames, with all the anticipated special relativistic effects (as exemplified in, e.g., \cite{Smith}).   The focus in this paper is on how the vacuum and thermal fluctuations in the quantum field affect the idf $\chi$ of a moving atom, specifically on how motion of the center of mass, the mdf,  affects its spread $\langle\chi^2\rangle$ used as a measure of the quantum dispersion at zero temperature  and quantum decoherence at zero and finite temperatures. 

\subsection{Main results and organization}

{
The primary analytical results from our fully relativistic treatment of motional decoherence in the internal state of a massive object moving at a constant velocity through a quantum field are presented in Sec.~\ref{S:oihtrt}, with a systematic comparison against the conventional approach, which starts with a non-relativistic calculation and then adds on the relativistic corrections, carried out in Sec.~\ref{S:ebtort}. The pivotal point is to recognize the fact that the  displacement variance of the internal oscillator is a Lorentz-invariant and use it as a diagnostic measure. From these analyses we establish three key physical results:
\begin{enumerate}
    \item \textit{Transient regime and time dilation:} Because the internal damping rate and oscillation frequency are intrinsic rest-frame quantities, a laboratory observer perceives the moving internal oscillator as effectively gaining mass  ($\gamma_{v}m$) and evolving more slowly. Consequently, relativistic time dilation delays both the relaxation rate and the onset of decoherence compared to ab initio nonrelativistic estimates.
    \item \textit{Zero-temperature vacuum invariance:} At zero temperature, in the asymptotic steady state, the fully covariant formulation predicts a displacement variance that is independent of the detector's velocity, faithfully preserving the Lorentz invariance of the Minkowski vacuum. By contrast, a hybrid model that modifies only the field phase via a leading-order Doppler shift erroneously predicts a decoherence measure that grows monotonically with velocity.
    \item \textit{Finite-temperature aberration:} When the ambient field is in a thermal state, which defines a preferred reference frame, the covariant noise spectrum is governed not only by Doppler shifts, but also by relativistic aberration. At high temperatures, this relativistic beaming concentrates hot incoming modes into a narrow forward cone while cooling the remaining angular directions, causing the steady-state decoherence measure to decrease at higher velocities — a non-monotonic feature completely missed by the conventional hybrid approximations.
\end{enumerate}
}

The central ``take-home'' message from our investigation can thus be summarized concisely: even when the internal degrees of freedom of a uniformly moving atom or detector do not oscillate at relativistic speeds in its rest frame, their dynamics must still abide by and be formulated in  accordance with the principles of special relativity.  Namely, the theoretical framework must intrinsically preserve the Lorentz-covariant structure of the combined atom-field interactions, consistently accounting for relativistic inertia scaling and mode aberration, regardless of whether the evolution is parameterized by proper time or laboratory coordinate time.  Adopting a hybrid model that couples a relativistic quantum field to nonrelativistic internal dynamics risks introducing spurious, symmetry-breaking artifacts that easily escape detection when a low-velocity expansion is imposed \textit{ab initio}.

This paper is organized as follows: In Sec.~\ref{S:yujyb}, we formulate the internal dynamics of a uniformly moving object within the laboratory frame, expressing the equations of motion and the displacement variance first in terms of the invariant proper time $\tau$ and subsequently in terms of the laboratory coordinate time $t$ to show the covariance property of internal dynamics. This is done for the specific purpose of contrasting with an ad-hoc nonrelativistic formulation in terms of $t$, which is a common practice. In Sec.~\ref{S:wittr}, we transition to the detector's comoving frame, deriving the exact finite-temperature Hadamard function and its noise power spectrum to analyze the behavior of the decoherence measure across both low- and high-temperature regimes. In Sec.~\ref{S:eotge}, for the purpose of comparison, we carry out a calculation using the conventional hybrid model, where the motion of the atom and its internal dynamics are nonrelativistic, but by adding corrections in the field, such as the relativistic Doppler effect. In Sec.~\ref{S:ebtort}, we uncover the exact origins of spurious artifacts in nonrelativistic treatments plus relativistic ``corrections'', and present a quantitative comparison that identifies such fallacies.

\section{Quantum Dispersion and Motional Decoherence}\label{S:yujyb}
\label{ch:motional-decoherence}

Because temperature is not a covariant quantity, a thermal field selects out a specific frame of reference.  In this section, we work with this frame, where the thermal fluctuations of the field appear homogeneous and isotropic, call it the lab frame.  We shall however describe the dynamics of the atom's internal degree of freedom in terms of its proper time $\tau$, as a proper relativistic treatment calls for. We subsequently recast the results into the lab time $t$ to underscore their Lorentz-invariant nature. In Section~\ref{S:eotge}, we will transition to the perspective of an observer comoving with the atom. These expressions provide a baseline for later comparisons, allowing us to pinpoint precisely where ad-hoc approximations with little or no regard for relativistic covariance may err.

\subsection{Internal dynamics described in proper-time $\tau$}

Consider a massive ($M$) object with internal degrees of freedom moving at constant velocity $\bm{v}$ in a massless quantum scalar field $\phi$. The internal state of the atom is modeled by a harmonic oscillator of mass $m$ and bare energy gap $\omega_{\textsc{b}}$, coupled linearly to the field along the atom's worldline $z^{\mu}(\tau)$ via the interaction action
\begin{equation}
    S_{\text{int}}=\lambda\int\!d\tau\; \chi(\tau)\,\phi(z^{\mu}(\tau))\,,
\end{equation}
where $\lambda$ denotes the coupling strength, $\tau$ is the Lorentz-invariant proper time, and $\chi(\tau)$ represents the displacement of the internal oscillator, which is a four-scalar by construction. Without loss of generality, this simple linearly coupled oscillator model offers an exactly solvable framework that enables us to pinpoint the theoretical pitfalls stemming from violations of Lorentz covariance or overly simplified kinematic assumptions.

For uniform motion parameterized by proper time $\tau$, the atom's worldline is given by
\begin{equation}
    z^{\mu}(\tau)=(\gamma_{v}\tau,\gamma_{v}\bm{v}\tau)=u^{\mu}\tau\,,
\end{equation}
where $\gamma_{v}=1/\sqrt{1-\bm{v}^2}$ is the Lorentz factor and $u^{\mu}=\gamma_{v}\,(1,\bm{v})$ is the four-velocity of the object, normalized by $u^{\mu}u_{\mu}=1$. Then, the equation of motion for the internal degree of freedom is given by varying the total action with respect to $\chi$,
\begin{equation}
    \ddot{\chi}(\tau)+\omega_{\textsc{b}}^2\chi(\tau)=\frac{\lambda}{m}\,\phi(z^{\mu}(\tau))\,,
\end{equation}
where an overdot denotes taking the derivative with respect to proper time $\tau$. The total field $\phi$ evaluated on the worldline consists of two contributions: the preexisting free field $\phi_{\text{h}}$ and the retarded radiated field $\phi_{\textsc{r}}$ generated by the atom's internal dof motion, i.e., 
\begin{equation}
    \phi_{\textsc{r}}(z^{\mu}(\tau))=\lambda\int_{0}^{\tau}\!d\tau'\;G_{\textsc{r}}^{(\phi)}(z^{\mu}(\tau),z^{\mu}(\tau'))\,\chi(\tau')\,.
\end{equation}  
In $(1+3)$-dimensional Minkowski spacetime, the retarded Green's function $G_{\textsc{r}}^{(\phi)}$ of the massless scalar field can be written in the manifestly covariant form
\begin{equation}
    G_{\textsc{r}}^{(\phi)}(z^{\mu}(\tau),z^{\mu}(\tau'))=\frac{1}{4\pi}\,\theta(\tau-\tau')\,\delta(\sigma(z^{\mu}(\tau),z^{\mu}(\tau')))\,,
\end{equation}
where $\sigma(x,x')$ is Synge's world function, which is half of the squared geodesic distance between the two spacetime points $x$ and $x'$. Along the inertial worldline $z^{\mu}(\tau)=u^{\mu}\tau$, the world function simplifies to
\begin{equation}
    \sigma(z^{\mu}(\tau),z^{\mu}(\tau'))=\frac{1}{2}\,\bigl(\tau-\tau'\bigr)^2\,.
\end{equation}
Then, after renormalization by absorbing the divergence in the retarded integral into the bare frequency and giving out a finite damping term, the equation of motion for the internal degree of freedom has the form:
\begin{equation}\label{E:fgnote}
    \ddot{\chi}(\tau)+2\gamma\,\dot{\chi}(\tau)+\omega^2\chi(\tau)=\frac{\lambda}{m}\,\phi_{\text{h}}(z^{\mu}(\tau))\,,
\end{equation}
where $\gamma=\lambda^2/(8\pi m)$ is the damping constant, and $\omega^2=\omega_{\textsc{b}}^2-\gamma^2$ is the squared renormalized frequency of the internal degree of freedom. The damping term $2\gamma\,\dot{\chi}(\tau)$ emerges from the backreaction of the retarded field $\phi_{\textsc{r}}$ on the internal dynamics,  manifested as radiation reaction. We will discuss the physical interpretation of this equation of motion in detail later.

The solution to this equation can be formally constructed using a fundamental set of homogeneous solutions, $d_{1}(\tau)$ and $d_{2}(\tau)$, of the damped harmonic oscillator, defined by the initial conditions $d_{1}(0)=1$, $\dot{d}_{1}(0)=0$, $d_{2}(0)=0$, and $\dot{d}_{2}(0)=1$. Then, the full solution $\chi$ can be written as 
\begin{equation}
    \chi(\tau)=\chi_{\text{h}}(\tau)+\frac{\lambda}{m}\int_{0}^{\tau}\!d\tau'\;d_{2}(\tau-\tau')\phi_{\text{h}}(z^{\mu}(\tau'))\,,
\end{equation}
where $\chi_{\text{h}}(\tau)$ is the homogeneous solution determined by the initial state of the oscillator,
\begin{equation}
    \chi_{\text{h}}(\tau)=d_{1}(\tau)\,\chi(0)+d_{2}(\tau)\,\dot{\chi}(0)\,,
\end{equation}
and the integral represents the inhomogeneous response driven by the free field $\phi_{\text{h}}$. The explicit forms of $d_{1}(\tau)$ and $d_{2}(\tau)$ are given by
\begin{align}\label{E:bgrtir}
    d_{1}(\tau)&=e^{-\gamma\tau}\,\biggl(\cos\Omega\tau+\frac{\gamma}{\Omega}\,\sin\Omega\tau\biggr)\,,&d_{2}(\tau)&=\frac{1}{\Omega}\,e^{-\gamma\tau}\sin\Omega\tau\,,
\end{align}
where $\Omega=\sqrt{\omega^2-\gamma^2}$ is the damped oscillation frequency of the internal degree of freedom. Here, causality is manifestly preserved in this formulation because the influence of the field at time $\tau'$ only affects the internal degree of freedom at later times $\tau>\tau'$.

\subsection{Measure of quantum dispersion and motional decoherence}

For a single atom, we will use $\langle\Delta\chi^{2}(\tau)\rangle=\langle\chi^{2}(\tau)\rangle-\langle\chi(\tau)\rangle^{2}$ to gauge the decoherence dynamics of the internal degree of freedom due to the atom's external motion. The expectation value $\langle\cdots\rangle$ is evaluated with respect to the initial state of the combined atom-field system. Before we delve into the details it may be helpful to first gain some intuitive understanding with a description of a simple and familiar case -- interference and diffraction patterns,  and then compare the measure we are using here with two commonly used measures -- purity and the von Neumann entropy -- in order to better appreciate its advantages.  

\subsubsection{A familiar example for intuitive understanding}
 
One of the most visually convincing demonstrations of the quantum property is to send the matter wave of particles, say electrons, through a pair of slits or a beamsplitter~\cite{Tonomura89}. On the screen behind the slits, we can observe an alternating bright and dark pattern. This is the interference pattern of the double-slit matter wave, in analogy to the famous Young's light interference pattern. In recent decades, similar experiments have been carried out using much larger particles, such as $\text{C}_{60}$, nanoparticle~\cite{ArndtAJP,Arndt2026}, or even   large organic molecules $\text{C}_{707}\text{H}_{260}\text{F}_{908}\text{N}_{16}\text{S}_{53}\text{Zn}_4$~\cite{Gerlich2011}.

If the particle interacts with the environment, such as the electron with the ambient stray quantized electromagnetic field, then the corresponding interference fringe tends to be fuzzy and less contrasting. This is a consequence of quantum decoherence, where, in the case of the electron, the Aharonov-Bohm phases of the two partial waves of the electron matter wave passing through the double slits experience different degrees of disruption caused by the quantum noises of the electromagnetic field~\cite{Ford93, Ford97, MazzitelliPaz2003, Levinson04, HsiangFord2004,HsiangLee06}. Thus, the phases of the partial waves arriving at the screen fluctuate. The coherence between the partial waves is lost. This randomly and rapidly shifting of the fringe pattern on the screen around its mean position, though in tiny amounts, causes it to smear out. The dark and white pattern appears subdued, losing fringe visibility. Degradation becomes more severe, if the environment is thermal. Large thermal fluctuations in a high-temperature electromagnetic field easily wash out the electron interference pattern, leaving two bright bands on the screen beyond the slits, each retaining its single-slit diffraction pattern. Similar effects are also observed for single-slit electron diffraction, where the fringes inside the diffraction envelope can get washed out due to electromagnetic field noise, and in the extreme case of decoherence, we only see a broad, featureless smear representing the classical sum of probabilities of the particle passing through different parts of the slit.

The decohering effect due to the quantum noise of the environment is also manifested in a single wavepacket of the system that interacts with the environment, where it causes the width of the wavepacket to be further spread out~\footnote{It can shrink the width as well, depending on the coherence properties of the environment, such as squeezing~\cite{HsiangFord2008}. Here we are only concerned with thermal noise.} than the intrinsic position dispersion of the system. There is a conceptual distinction between quantum position dispersion and environment-induced broadening. The latter is irreversible~\cite{HsiangHu22_NoIntrinsic} since phase information embedded in the wavepacket of the system is partially lost. Heuristically, the environment quantum noise drives the wavepacket along a randomly zigzag path, introducing smearing to the probability distribution, described by  the wavepacket's position. Thus, we may characterize increasing position dispersion as a physically meaningful indicator of the tendency toward classicality, where the classical statistical distribution arises~\footnote{Here we would like to emphasize an important distinction~\cite{HsiangHu22_NoIntrinsic} between two often confused pictures of classicality: a classical point particle and a classical statistical distribution. Ideally a classical point particle is described by a delta function, but in practice we may not have knowledge of its precise location except for a classical statistical distribution.}

\subsubsection{Comparison with purity and von Neumann entropy}

Quantitatively, let us compare the environment-induced decoherence measure used here with two commonly used decoherence measures for a Gaussian system, namely,  purity $\mu$ and von Neumann entropy $S_{\text{vN}}$, defined  respectively by
\begin{align}
    \mu&=\operatorname{Tr}\bigl(\rho^2\bigr)=\frac{1}{2\sqrt{u}}\,,\\
    S_{\text{vN}}&=\operatorname{Tr}\bigl(\rho\,\ln\rho\bigr)=\Bigl(\sqrt{u}+\frac{1}{2}\Bigr)\,\ln\Bigl(\sqrt{u}+\frac{1}{2}\Bigr)-\Bigl(\sqrt{u}-\frac{1}{2}\Bigr)\,\ln\Bigl(\sqrt{u}-\frac{1}{2}\Bigr)\,,
\end{align}
where $u$ is the uncertainty function, given by the determinant of the covariance matrix element $V_{ij}=\dfrac{1}{2}\operatorname{Tr}\bigl(\rho\,\bigl\{Z,\,Z^T\bigr\}\bigr)$. The column vector $Z$ contains the dynamical variables $(\chi,p)$ of the reduced Gaussian system in a state described by the density matrix operator $\rho$. The variable $p$ is the canonical momentum conjugate to $\chi$. Here we only use the position uncertainty/variance, one element of the covariance matrix, as the decoherence measure. In general, this does not provide complete information, but in the present setting, it serves as a highly effective and physically intuitive diagnostic.

First, from the perspective of a which-path experiment, a spatially spreading wavepacket provides greater discernibility for the environment to distinguish between widely separated parts of the wavepacket, naturally increasing the system's susceptibility to decoherence. Second, the calculation of position variance is well-defined and ultraviolet finite after frequency renormalization. In contrast, calculating purity requires the momentum uncertainty, which is ultraviolet-sensitive and necessitates a high-frequency cutoff. Introducing such a cutoff scale can be ambiguous and risks inadvertently violating Lorentz invariance, the exact theoretical pitfall we aim to avoid. Finally, position uncertainty can still quantitatively track purity in the appropriate limits. As shown later, relativistic time dilation suppresses the effective damping rate, acting effectively as a weaker coupling. In this weak-coupling limit, cutoff-dependent contributions become negligible, and the asymptotic equilibrium state of the Gaussian system is approximately thermal and unsqueezed~\cite{HsiangHu2021}. Under these conditions, the momentum uncertainty becomes proportional to the position uncertainty, 
\begin{align*}
    \langle \Delta p^2\rangle&=m^2\omega^2\,\langle \Delta \chi^2\rangle\,,&&\text{with}&\langle\bigl\{\Delta x,\,\Delta p\bigr\}\rangle&=0\,,
\end{align*}
making the system's purity inversely proportional to its position variance 
\begin{equation}
    \mu=\frac{1}{2\sqrt{\langle \Delta \chi^2\rangle\,\langle \Delta p^2\rangle-\dfrac{1}{4}\langle\bigl\{\Delta x,\,\Delta p\bigr\}\rangle}}=\frac{1}{2m\omega\,\langle \Delta \chi^2\rangle}\,.
\end{equation}
Taking these factors into consideration, the position uncertainty serves as a physically meaningful, covariant proxy for quantifying asymptotic purity, i.e., a decoherence measure for the current setting. In particular, the position uncertainty grows increasingly larger with higher environmental temperature $\beta^{-1}$. For example, the position uncertainty of a quantum harmonic oscillator coupled to a massless quantum scalar field in the limit of vanishing coupling will approach this value
\begin{equation}
    \langle \Delta \chi^2\rangle\simeq\frac{1}{2m\omega}\,\coth\frac{\beta\omega}{2}\,,
\end{equation}
in the asymptotic equilibrium state. At very high temperature $\beta\to0$, it approaches the classical value
\begin{equation}
    \langle \Delta \chi^2\rangle\simeq\frac{1}{2m\omega^2\beta}\,.
\end{equation}
We now return to the technical development. 

\subsubsection{Technicalities of the measure used here}

 To preclude nonessential technical complications, we assume a factorized initial state, $\hat{\rho}(0)=\hat{\rho}_{\chi}(0)\otimes\hat{\rho}_{\phi}(0)$, where $\hat{\rho}_{\chi}(0)$ and $\hat{\rho}_{\phi}(0)$ are the initial density matrices of the internal degree of freedom and the field, respectively. Suppose the initial state has zero-mean, uncorrelated initial conditions, and $\langle\chi(0)\rangle=\langle\dot{\chi}(0)\rangle=0$, $\langle\{\chi(0),\dot{\chi}(0)\}\rangle=0$ alongside $\langle\phi_{\text{h}}(0)\rangle=0$. Formally, the variance takes the form,
\begin{align}
    \langle\Delta\chi^{2}(\tau)\rangle=\langle\chi^{2}(\tau)\rangle&=d_{1}^{2}(\tau)\,\langle\chi^{2}(0)\rangle+d_{2}^{2}(\tau)\,\langle\dot{\chi}^{2}(0)\rangle\notag\\
    &\qquad\qquad+\frac{\lambda^{2}}{m^{2}}\int_{0}^{\tau}\!d\tau'\,d\tau''\;d_{2}(\tau-\tau')\,d_{2}(\tau-\tau'')\,G_{\textsc{h}}^{(\phi)}(z^{\mu}(\tau'),z^{\mu}(\tau''))\,.
\end{align}
The uncertainty in the displacement of the internal degree of freedom characterizes the spatial width of the wave packet of the internal degree of freedom. When the atom is decoupled from the environmental quantum field, i.e., $\lambda=0$, this quantity reduces purely to the first two terms with $\gamma=0$, capturing the intrinsic \textit{quantum dispersion} of the isolated oscillator, usually illustrated in textbooks by the spread of a wavepacket. In this decoupled limit, the internal subsystem is closed, its evolution remains unitary, and no decoherence occurs.

Turning on the interaction, $\lambda\neq0$, subjects the internal oscillator to both radiative damping, encoded in the exponential decay of $d_1(\tau)$ and $d_2(\tau)$, and stochastic quantum field fluctuations, governed by the double integral over the Hadamard function $G_{\textsc{h}}^{(\phi)}$. These environmental quantum noises drive the reduced state of the atom into a statistical mixture, rendering its evolution non-unitary and degrading its internal quantum coherence. External motion of the atom further compounds this process in a nontrivial manner, giving rise to  \textit{motional decoherence}. Since environmental decoherence rates generically scale with the squared coordinate separation $\lvert\chi-\chi'\rvert^2$ in the reduced density matrix, a broader spatial spread $\langle\Delta\chi^2(\tau)\rangle$ exposes more widely separated locations in the internal wave packet to greater distinguishability by the environment through their interactions with the quantum field, signifying the loss of spatial coherence.

The Hadamard function of the massless scalar field, which serves as the noise kernel $G_{\textsc{h}}^{(\phi)}$, is defined along the worldline by
\begin{equation}
    G_{\textsc{h}}^{(\phi)}(z^{\mu}(\tau'),z^{\mu}(\tau''))=\frac{1}{2}\,\langle\bigl\{\phi_{\text{h}}(z^{\mu}(\tau')),\phi_{\text{h}}(z^{\mu}(\tau''))\bigr\}\rangle\,,
\end{equation}
and encodes the spacetime correlation of the free-field fluctuations. From an open-system perspective, it characterizes the statistical strength of the quantum noise from the field environment. Whenever the field fluctuations are stationary along the atom's worldline, meaning, that the correlation function is invariant under shifts in proper time, the Hadamard function depends only on the difference of the proper times, i.e., $G_{\textsc{h}}^{(\phi)}(z^{\mu}(\tau'),z^{\mu}(\tau''))=G_{\textsc{h}}^{(\phi)}(\tau'-\tau'')$. In this case, we can express the decoherence measure as
\begin{align}\label{E:dfgeir}
    \langle\Delta\chi^{2}(\tau)\rangle&=d_{1}^{2}(\tau)\,\langle\chi^{2}(0)\rangle+d_{2}^{2}(\tau)\,\langle\dot{\chi}^{2}(0)\rangle+\frac{\lambda^{2}}{m^{2}}\int_{0}^{\tau}\!d\tau'\,d\tau''\;d_{2}(\tau')\,d_{2}(\tau'')\,G_{\textsc{h}}^{(\phi)}(\tau'-\tau'')\,.
\end{align}
The frame independence of this decoherence measure follows directly from its construction as a Lorentz scalar. After we recast the equation of motion and its solution in terms of the lab time $t$, we can clearly show the manifestly invariant structure.

For uniform motion at constant velocity $\bm{v}$, the lab time interval $dt$ and the proper time interval $d\tau$ are related simply by $dt=\gamma_{v}\,d\tau$. The proper-time equation of motion, Eq.~\eqref{E:fgnote}, can then be recast in terms of the lab time $t$ as
\begin{equation}\label{E:rnpfg}
    \chi''(t)+2\frac{\gamma}{\gamma_{v}}\,\chi'(t)+\frac{\omega^2}{\gamma_{v}^2}\,\chi(t)=\frac{\lambda}{m\,\gamma^{2}_{v}}\,\phi_{\text{h}}(z^{\mu}(t))\,,
\end{equation}
where a prime denotes differentiation with respect to the lab time $t$. Its solution is given by
\begin{equation}
    \chi(t)=\mathsf{d}_{1}(t)\,\chi(0)+\mathsf{d}_{2}(t)\,\chi'(0)+\frac{\lambda}{m\,\gamma^{2}_{v}}\int_{0}^{t}\!dt'\;\mathsf{d}_{2}(t-t')\,\phi_{\text{h}}(z^{\mu}(t'))
\end{equation}
where the fundamental solutions in lab-frame are related to their proper-time counterparts via $\mathsf{d}_{1}(t)=d_{1}(t/\gamma_{v})$ and $\mathsf{d}_{2}(t)=\gamma_{v}\,d_{2}(t/\gamma_{v})$. In terms of $t$, the decoherence measure evaluates to   
\begin{align}\label{E:ngldgj}
    \langle\Delta\chi^{2}(t)\rangle&=\mathsf{d}_{1}^{2}(t)\,\langle\chi^{2}(0)\rangle+\mathsf{d}_{2}^{2}(t)\,\langle\chi'^{2}(0)\rangle+\frac{\lambda^{2}}{m^{2}\gamma^{4}_{v}}\int_{0}^{t}\!dt'\,dt''\;\mathsf{d}_{2}(t')\,\mathsf{d}_{2}(t'')\,G_{\textsc{h}}^{(\phi)}(t'-t'')\,.
\end{align}
By applying the scaling relations $dt=\gamma_{v}\,d\tau$ and $\chi'(0)=\dot{\chi}(0)/\gamma_v$, we can directly verify that Eq.~\eqref{E:ngldgj} is algebraically identical to Eq.~\eqref{E:dfgeir}, explicitly confirming the reparameterization and Lorentz-invariant nature of the displacement variance.

Inspection of Eq.~\eqref{E:rnpfg} also reveals how the atom's external motion modifies the apparent parameters of its internal dynamics. Both the effective damping rate $\gamma/\gamma_v$ and the effective oscillation frequency $\omega/\gamma_v$ are suppressed by the Lorentz factor $\gamma_v$, reflecting the fact that \textit{$\gamma$ and $\omega$ are intrinsic quantities defined in the atom's rest frame}. This suppression is a direct manifestation of relativistic time dilation, which slows down the internal evolution as observed from the lab frame, a feature equally evident in the dilated time arguments of $\mathsf{d}_i(t)$. Meanwhile, the effective mass of the oscillator increases by the same  factor, $m\to m\,\gamma_v$, while the effective coupling to the field is reduced to $\lambda/\gamma_v$, together accounting for the $\gamma_v^{-2}$ suppression of the driven contribution on the right-hand side of Eq.~\eqref{E:rnpfg}. Effectively, a lab observer perceives a more massive internal oscillator that evolves and dissipates energy more slowly. Consequently, in lab time, the internal degree of freedom requires a longer duration to relax to equilibrium, and its decoherence rate is correspondingly slower.

Note that so far we have remained within the lab frame, merely expressing the internal dynamics in terms of two different temporal parameters: the invariant proper time $\tau$ and the lab time $t$. We now shift our perspective to that of an observer comoving with the atom.

\section{Internal dynamics described by an observer comoving with the atom}\label{S:wittr}

In the comoving frame, parameterized by the proper time $\tau$ with the atom fixed at the spatial origin $\bm{x}=0$, the atom is at rest and perceives the quantum field as moving with velocity $-\bm{v}$. In this frame, the internal dynamics is governed directly by the proper-time equation of motion, Eq.~\eqref{E:fgnote}, along with the fundamental solutions $d_i(\tau)$ given in Eq.~\eqref{E:bgrtir}. Our primary task here is  to evaluate the explicit form of the field's Hadamard function, $G_{\textsc{h}}^{(\phi)}(\tau'-\tau'')=\dfrac{1}{2}\,\langle\bigl\{\phi_{\text{h}}(z^{\mu}(\tau')),\,\phi_{\text{h}}(z^{\mu}(\tau''))\bigr\}\rangle$, which governs the decoherence dynamics of the internal degree of freedom. Although the field anti-commutator $\bigl\{\phi_{\text{h}}(z^{\mu}(\tau')),\,\phi_{\text{h}}(z^{\mu}(\tau''))\bigr\}$ is a Lorentz scalar operator by construction, its expectation value, the Hadamard function, depends on the initial state of the field. While evaluating this expectation value in the laboratory or comoving coordinates must yield the same physical result for a given trajectory and bath configuration, whether the resulting noise kernel is Lorentz invariant, i.e., independent of the detector's velocity 
$\bm{v}$, depends entirely on whether the initial field state itself respects Lorentz symmetry.

\subsection{Hadamard function of a quantum field at finite temperature}

When the initial state of the quantum field is prepared as a thermal equilibrium state at an inverse temperature $\beta$, the Hadamard function evaluated along the atom's uniform trajectory $\bm{x}(t)=\bm{v}\,t$ in the lab frame is given by
\begin{align}
    G_{\textsc{h}}^{(\phi)}(t,\bm{x};t',\bm{x}')&=\int\!\frac{d^{3}\bm{\mathsf{k}}}{(2\pi)^{3}}\frac{1}{2w_{\bm{\mathsf{k}}}}\;\biggl(N_{w_{\bm{\mathsf{k}}}}+\frac{1}{2}\biggr)\,\biggl[e^{-i(w_{\bm{\mathsf{k}}}-\bm{\mathsf{k}}\cdot\bm{v})(t-t')}+\text{C.C.}\biggr]\,,
\end{align}
where $w_{\bm{\mathsf{k}}}$ denotes the frequency of mode $\bm{\mathsf{k}}$ measured in the lab frame, and
\begin{equation*}
    N_{w_{\bm{\mathsf{k}}}}+\frac{1}{2}=\frac{1}{2}\,\coth\frac{\beta w_{\bm{\mathsf{k}}}}{2}\,.
\end{equation*}
Substituting this distribution factor explicitly yields
\begin{align}
    G_{\textsc{h}}^{(\phi)}(t,\bm{x};t',\bm{x}')&=\int\!\frac{d^{3}\bm{\mathsf{k}}}{(2\pi)^{3}}\frac{1}{4w_{\bm{\mathsf{k}}}}\;\coth\frac{\beta w_{\bm{\mathsf{k}}}}{2}\,\biggl[e^{-i(w_{\bm{\mathsf{k}}}-\bm{\mathsf{k}}\cdot\bm{v})(t-t')}+\text{C.C.}\biggr]\label{E:bgeit}\\ 
    &=\int\!\frac{d^{3}\bm{\mathsf{k}}}{(2\pi)^{3}}\frac{1}{4w_{\bm{\mathsf{k}}}}\;\coth\frac{\beta w_{\bm{\mathsf{k}}}}{2}\,\biggl[e^{-i\gamma_{v}(w_{\bm{\mathsf{k}}}-\bm{\mathsf{k}}\cdot\bm{v})(\tau-\tau')}+\text{C.C.}\biggr]\,,\label{E:dbetje}
\end{align}
where in the last line we have used the relation $t-t'=\gamma_{v}\,(\tau-\tau')$ connecting the lab time interval and the proper time interval along the atom's worldline. These two equivalent forms represent the thermal noise kernel evaluated entirely within the lab frame, differing solely in their choice of temporal parameterization.

To evaluate the Hadamard function from the perspective of the atom's comoving frame, it is more convenient to parameterize the momentum space using the wave vector $\bm{k}$ and the corresponding frequency $\omega_{\bm{k}}$ measured in the rest frame of the atom. These variables are related to their lab-frame counterparts $\bm{\mathsf{k}}$ and $w_{\bm{\mathsf{k}}}=\lvert\bm{\mathsf{k}}\rvert$ by a Lorentz transformation,
\begin{align}
    \omega_{\bm{k}}=\lvert\bm{k}\rvert&=\gamma_{v}\,\bigl(w_{\bm{\mathsf{k}}}-\bm{k}\cdot\bm{v}\bigr)\,,&\bm{k}&=\bm{\mathsf{k}}+\frac{\gamma_{v}-1}{v^2}\,(\bm{\mathsf{k}}\cdot\bm{v})\,\bm{v}-\gamma_{v}\,w_{\bm{\mathsf{k}}}\,\bm{v}\,.   
\end{align}
By exploiting the Lorentz invariance of the momentum-space integration measure,
\begin{equation*}
    \frac{d^{3}\bm{k}}{2\omega_{\bm{k}}}=\frac{d^{3}\bm{\mathsf{k}}}{2w_{\bm{\mathsf{k}}}}\,,
\end{equation*}
the Hadamard function $G_{\textsc{h}}^{(\phi)}(\tau'-\tau'')$ in Eq.~\eqref{E:dfgeir} can be expressed in terms of comoving variables as
\begin{align}\label{E:dbgrt}
    G_{\textsc{h}}^{(\phi)}(\tau'-\tau'')&=\int\!\frac{d^{3}\bm{k}}{(2\pi)^{3}}\frac{1}{4\omega_{\bm{k}}}\;\coth\frac{\beta w_{\bm{\mathsf{k}}}}{2}\,\biggl[e^{-i\omega_{\bm{k}}(\tau'-\tau'')}+\text{C.C.}\biggr]\notag\\
    &=\int\!\frac{d^{3}\bm{k}}{(2\pi)^{3}}\frac{1}{4\omega_{\bm{k}}}\;\coth\frac{\beta\omega_{\bm{k}}}{2\gamma_{v}(1-v\,s)}\,\biggl[e^{-i\omega_{\bm{k}}(\tau'-\tau'')}+\text{C.C.}\biggr]\,,
\end{align}
where $s=\cos\theta$, with $\theta$ being the angle between the lab-frame momentum $\bm{\mathsf{k}}$ and the atom's velocity $\bm{v}$. Alternatively, it is more convenient to evaluate this using the angle $\vartheta$ between the comoving-frame momentum $\bm{k}$ and $\bm{v}$. These two angles are linked by the relativistic aberration formulas,
\begin{align}\label{E:neotg}
    \varsigma&=\cos\vartheta=\frac{s-v}{1-v\,s}\,,&s&=\frac{\varsigma+v}{1+v\,\varsigma}\,,
\end{align}
which yields the identity
\begin{equation}
    \frac{1}{1-v\,s}=\gamma_{v}^{2}\,(1+v\,\varsigma)\,.
\end{equation}
Consequently, the argument of the hyperbolic cotangent function can be rewritten entirely in terms of the comoving-frame variables as
\begin{equation}\label{E:dbgrtz}
    \frac{\beta w_{\bm{\mathsf{k}}}}{2}=\frac{\beta\omega_{\bm{k}}}{2\gamma_{v}(1-v\,s)}=\frac{\beta\omega_{\bm{k}}}{2}\,\gamma_{v}\,(1+v\,\varsigma)\,,
\end{equation}
and the Hadamard function  expressed in terms of the proper time is given by
\begin{align}
    G_{\textsc{h}}^{(\phi)}(\tau'-\tau'')&=\int\!\frac{d^{3}\bm{k}}{(2\pi)^{3}}\frac{1}{4\omega_{\bm{k}}}\;\coth\frac{\beta\omega_{\bm{k}}\gamma_{v}(1+v\,\varsigma)}{2}\,\biggl[e^{-i\omega_{\bm{k}}(\tau'-\tau'')}+\text{C.C.}\biggr]\,.
\end{align}
Since the Hadamard function $G_{\textsc{h}}^{(\phi)}(\tau'-\tau'')$ is a Lorentz invariant scalar by construction, its physical value remains independent of the choice of reference frame, a property explicitly verified by the preceding coordinate transformations. Formulating the noise kernel in terms of proper time provides an elegant framework for analyzing quantum decoherence, particularly in vacuum states where exact Lorentz covariance is manifestly preserved.

Although Eqs.~\eqref{E:dbetje} and \eqref{E:dbgrt} appear superficially similar, they should not be identified term by term. In Eq.~\eqref{E:dbetje}, the field modes are parametrized in the lab frame, so the thermal factor has the standard isotropic form, while the phase contains the Doppler shift due to the atom's motion. In Eq.~\eqref{E:dbgrt}, the modes are parametrized in the atom's rest frame, where the phase simplifies, but the thermal factor becomes anisotropic because the bath is moving in that frame. This crucial distinction is easily overlooked if the two expressions are compared solely at a formal level.

Mathematically, the four-wave vectors $(w_{\bm{\mathsf{k}}},\bm{\mathsf{k}})$ and $(\omega_{\bm{k}},\bm{k})$ denote the same physical field mode as measured in different inertial frames and are related by a Lorentz transformation. The two forms of the Hadamard function are therefore physically equivalent, but the effects of the relative motion are encoded in different factors.

To make precise the physical meaning of these frame-dependent frequencies, we can express the frequency of the mode $\bm{k}^{\mu}$ measured by any observer with a four-velocity $u^{\mu}$ as the invariant contraction
\begin{equation}
    \omega_{\text{obs}}=k_{\mu}u^{\mu}\,.
\end{equation}
Thus, for an observer comoving with the atom, she has $u^{\mu}_{\text{atom}}=\gamma\,(1,\bm{v})$. Then for a massless lab-frame mode $\mathsf{k}^{\mu}=(w_{\bm{\mathsf{k}}},\bm{\mathsf{k}})$, she finds
\begin{equation}
    \omega_{\text{com}}=k_{\mu}u^{\mu}_{\text{atom}}=\gamma_{v}\,(w_{\bm{\mathsf{k}}}-\bm{\mathsf{k}}\cdot\bm{v})=\omega_{\bm{k}}\,.
\end{equation} 
On the other hand, for an observer at rest in the lab frame, he has $u^{\mu}_{\text{lab}}=(1,\bm{0})$. Then he finds
\begin{equation}
    \omega_{\text{lab}}=k_{\mu}u^{\mu}_{\text{lab}}=w_{\bm{\mathsf{k}}}\,,
\end{equation}
entirely consistent with the above discussion.

\subsection{Spectral distribution of field fluctuations}
The Fourier transform of the Hadamard function defines the noise power spectrum or spectral density experienced along the atom's trajectory, characterizing how environmental quantum fluctuations are distributed across frequency. This frequency-space perspective is physically essential for determining how different field modes resonantly drive the atom, governing the relative rates of absorption and emission. From a computational standpoint, transitioning to frequency space is equally vital. It transforms cumbersome double time-integrals into algebraic expressions, rendering the numerical evaluation of the internal degree of freedom's covariance matrix in the late-time equilibrium regime both tractable and computationally efficient.

With respect to the lab-frame time, the Fourier transform of the Hadamard function is given by
\begin{align}\label{E:gnortr}
    \widetilde{G}_{\textsc{h}}^{(\phi)}(\kappa)=\int_{-\infty}^{\infty}\!d\Delta t\; G_{\textsc{h}}^{(\phi)}(\Delta)\,e^{i\kappa\Delta t}
    &=\int\!\frac{d^{3}\bm{\mathsf{k}}}{(2\pi)^{2}}\;\frac{1}{4w_{\bm{\mathsf{k}}}}\,\coth\frac{\beta w_{\bm{\mathsf{k}}}}{2}\,\Bigl[\delta(\kappa-w_{\bm{\mathsf{k}}}+\bm{\mathsf{k}}\cdot\bm{v})+\delta(\kappa+w_{\bm{\mathsf{k}}}-\bm{\mathsf{k}}\cdot\bm{v})\Bigr]\notag\\ 
    &=\frac{1}{8\pi v}\int_{\frac{\lvert\kappa\rvert}{1+v}}^{\frac{\lvert\kappa\rvert}{1-v}}\!dw_{\bm{\mathsf{k}}}\;\coth\frac{\beta w_{\bm{\mathsf{k}}}}{2}\notag\\
    &=\frac{1}{4\pi \beta v}\,\ln\frac{\sinh\dfrac{\beta\lvert\kappa\rvert}{2(1-v)}}{\sinh\dfrac{\beta\lvert\kappa\rvert}{2(1+v)}}\,,
\end{align}
where $\Delta t=t'-t''$. The result in Eq.~\eqref{E:gnortr} can be further decomposed into two distinct physical contributions,
\begin{align}
    \widetilde{G}_{\textsc{h}}^{(\phi)}(\kappa)=\frac{\lvert\kappa\rvert}{4\pi}\,\gamma^{2}_{v}+\frac{1}{4\pi \beta v}\,\ln\frac{1-e^{-\frac{\beta\lvert\kappa\rvert}{1-v}}}{1-e^{-\frac{\beta\lvert\kappa\rvert}{1+v}}}\,.
\end{align}
The first term represents the temperature-independent vacuum contribution, while the second term accounts for the finite-temperature thermal environment.

We can carry out an analogous transformation for the proper-time Hadamard function in the atom's comoving frame, given by Eq.~\eqref{E:dbgrtz}. Its Fourier transform reads
\begin{align}\label{E:owindgf}
    \widehat{G}_{\textsc{h}}^{(\phi)}(\varkappa)&=\int_{-\infty}^{\infty}\!d\Delta\tau\; G_{\textsc{h}}^{(\phi)}(\Delta\tau)\,e^{i\varkappa\Delta\tau}\notag\\
    &=\int\!\frac{d^{3}\bm{k}}{(2\pi)^{2}}\;\frac{1}{4\omega_{\bm{k}}}\,\coth\frac{\beta\omega_{\bm{k}}\gamma_{v}(1+v\,\varsigma)}{2}\,\Bigl[\delta(\varkappa-\omega_{\bm{k}})+\delta(\varkappa+\omega_{\bm{k}})\Bigr]\notag\\
    &=\frac{\lvert\varkappa\rvert}{8\pi}\int_{-1}^{1}\!d\varsigma\;\coth\frac{\beta\lvert\varkappa\rvert\gamma_{v}(1+v\,\varsigma)}{2}\notag\\
    &=\frac{1}{4\pi\beta\gamma_{v}v}\,\ln\frac{\sinh\dfrac{\beta\lvert\varkappa\rvert}{2\gamma_{v}(1-v)}}{\sinh\dfrac{\beta\lvert\varkappa\rvert}{2\gamma_{v}(1+v)}}\,.
\end{align}
Here $\Delta\tau=\tau'-\tau''$ is the proper time interval, and $\varkappa$ denotes the frequency measured in the rest frame of the atom. Since the transform is taken with respect to the atom's proper-time interval, $\widehat G_H^{(\phi)}(\varkappa)$ provides a frame-independent description of the spectrum along the atom's worldline. Its dependence on $v$ reflects the physical relative motion between the atom and the thermal bath.

In contrast, the lab-frame $\widetilde{G}_{\textsc{h}}^{(\phi)}(\kappa)$ in Eq.~\eqref{E:gnortr} is frame-dependent. However, both are related by
\begin{equation}
    \widehat{G}_{\textsc{h}}^{(\phi)}(\varkappa)=\frac{1}{\gamma_{v}}\,\widetilde{G}_{\textsc{h}}^{(\phi)}(\frac{\varkappa}{\gamma_{v}})\,.
\end{equation}
We can likewise decompose $\widehat{G}_{\textsc{h}}^{(\phi)}(\varkappa)$ into two physically distinct parts,
\begin{align}\label{E:yiyjfg}
    \widehat{G}_{\textsc{h}}^{(\phi)}(\varkappa)=\frac{\lvert\varkappa\rvert}{4\pi}+\frac{1}{4\pi\beta\gamma_{v}v}\,\ln\frac{1-e^{-\frac{\beta\lvert\varkappa\rvert}{\gamma_{v}(1-v)}}}{1-e^{-\frac{\beta\lvert\varkappa\rvert}{\gamma_{v}(1+v)}}}\,.
\end{align}
The vacuum contribution takes the familiar invariant form, $\dfrac{\lvert\varkappa\rvert}{4\pi}$, and is independent of the atom's velocity. This feature reflects the underlying Lorentz invariance of the Minkowski vacuum. When the noise spectrum is properly formulated in terms of the atom's proper time, uniform motion does not alter the vacuum fluctuations registered by the atom.

\subsection{Motional decoherence by vacuum and thermal fluctuations}\label{S:oihtrt}

In terms of either Fourier transform of the Hadamard functions, we can greatly simplify the calculations of $\langle\Delta\chi^{2}\rangle$ from a double time integral to a single frequency integral, thereby rendering numerical evaluations remarkably efficient. For example, using the Fourier transform $\widehat{G}_{\textsc{h}}^{(\phi)}(\varkappa)$ in Eq.~\eqref{E:yiyjfg}, we have the induced or passive contribution to the decoherence measure $\langle\Delta\chi^{2}(\tau)\rangle$ cast as
\begin{align}
    \langle\Delta\chi^{2}(\tau)\rangle_{\text{passive}}&=\frac{\lambda^{2}}{m^{2}}\int_{0}^{\tau}\!d\tau'\,d\tau''\;d_{2}(\tau')\,d_{2}(\tau'')\int_{-\infty}^{\infty}\!\frac{d\varkappa}{2\pi}\;\widehat{G}_{\textsc{h}}^{(\phi)}(\varkappa)\,e^{-i\varkappa(\tau'-\tau'')}\notag\\ 
    &=\frac{\lambda^{2}}{m^{2}}\int_{-\infty}^{\infty}\!\frac{d\varkappa}{2\pi}\;\widehat{G}_{\textsc{h}}^{(\phi)}(\varkappa)\,\lvert\widehat{d}_{2}(\varkappa;\tau)\rvert^{2}\,,
\end{align}
where 
\begin{align}
    \widehat{d}_{2}(\varkappa;\tau)&=\int_{0}^{\tau}\!d\tau'\;d_{2}(\tau')\,e^{i\varkappa\tau'}=\frac{1}{i\,2\Omega}\biggl[\frac{1-e^{-[\gamma-i(\varkappa+\Omega)]\tau}}{\gamma-i(\varkappa+\Omega)}-\frac{1-e^{-[\gamma+i(\varkappa-\Omega)]\tau}}{\gamma+i(\varkappa-\Omega)}\biggr]\,,
\end{align}
such that
\begin{align}
    &\quad\lvert\widehat{d}_{2}(\varkappa;\tau)\rvert^{2}\notag\\&=\frac{1}{[\gamma^2+(\varkappa+\Omega)^2][\gamma^2+(\varkappa-\Omega)^2]}\notag\\
    &\qquad\qquad\qquad\times\Biggl\{1-\frac{2}{\Omega}\,e^{-\gamma\tau}\Bigl[\varkappa\,\sin\varkappa\tau\,\cos\Omega\tau+\Omega\,\cos\varkappa\tau\,\cos\Omega\tau+\gamma\,\cos\varkappa\tau\,\sin\Omega\tau\Bigr]\Biggr.\notag\\ 
    &\qquad\qquad\qquad+\Biggl.\frac{1}{4\Omega^{2}}\,e^{-2\gamma\tau}\,\Bigl[\bigl(\varkappa+\Omega^{2}+\gamma^{2}\bigr)-\bigl(\varkappa-\Omega^{2}+\gamma^{2}\bigr)\,\cos2\Omega\tau+2\Omega\gamma\,\sin2\Omega\tau\Bigr]\Biggr\}\,.
\end{align}
In particular, when the internal dynamics reaches its steady-state equilibrium in the asymptotic limit $\tau\to\infty$, $\lvert\widehat{d}_{2}(\varkappa;\tau)\rvert^{2}$ has a very simple form,
\begin{equation}
    \lim_{\tau\to\infty}\lvert\widehat{d}_{2}(\varkappa;\tau)\rvert^{2}=\frac{1}{[\gamma^2+(\varkappa+\Omega)^2][\gamma^2+(\varkappa-\Omega)^2]}\,.
\end{equation}
Meanwhile, the remaining transient contributions associated with the initial state rapidly decay to zero due to the presence of radiative damping.

The expression $\lvert\widehat{d}_{2}(\varkappa;\tau)\rvert^{2}$ is independent of the inverse temperature $\beta$ and the atom velocity $v$ since it depends exclusively on the intrinsic parameters of the internal oscillator. Consequently, any parameter dependence in the decoherence measure is entirely governed by the environmental noise kernel, $\widehat{G}_{\textsc{h}}^{(\phi)}(\varkappa)$. In the zero-temperature limit $\beta\to\infty$, the Hadamard function becomes independent of $v$, reflecting the manifest Lorentz invariance of the Minkowski vacuum. Thus, \textit{At zero temperature the decoherence measure is independent of the atom velocity}, establishing that uniform motion through the quantum vacuum does not alter the decoherence dynamics of \textit{the internal degrees of freedom}. 

\begin{figure}[htpb]
    \centering
    \begin{subfigure}{0.45\linewidth}
        \centering
        \includegraphics[width=\linewidth]{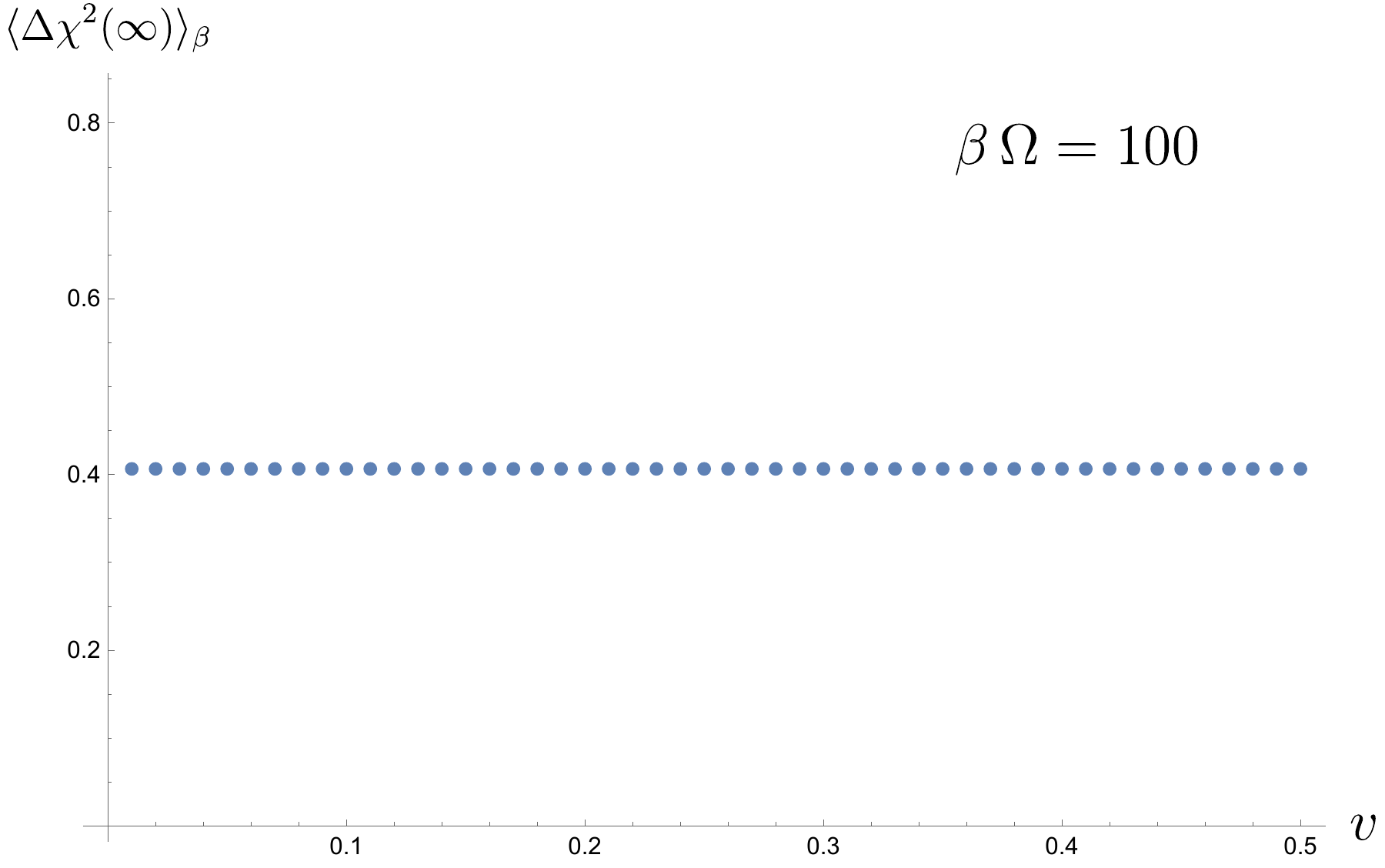}
        \caption{$\beta\Omega=100$}
        \label{Fi:graph1}
    \end{subfigure}
    \hfill
    \begin{subfigure}{0.45\linewidth}
        \centering
        \includegraphics[width=\linewidth]{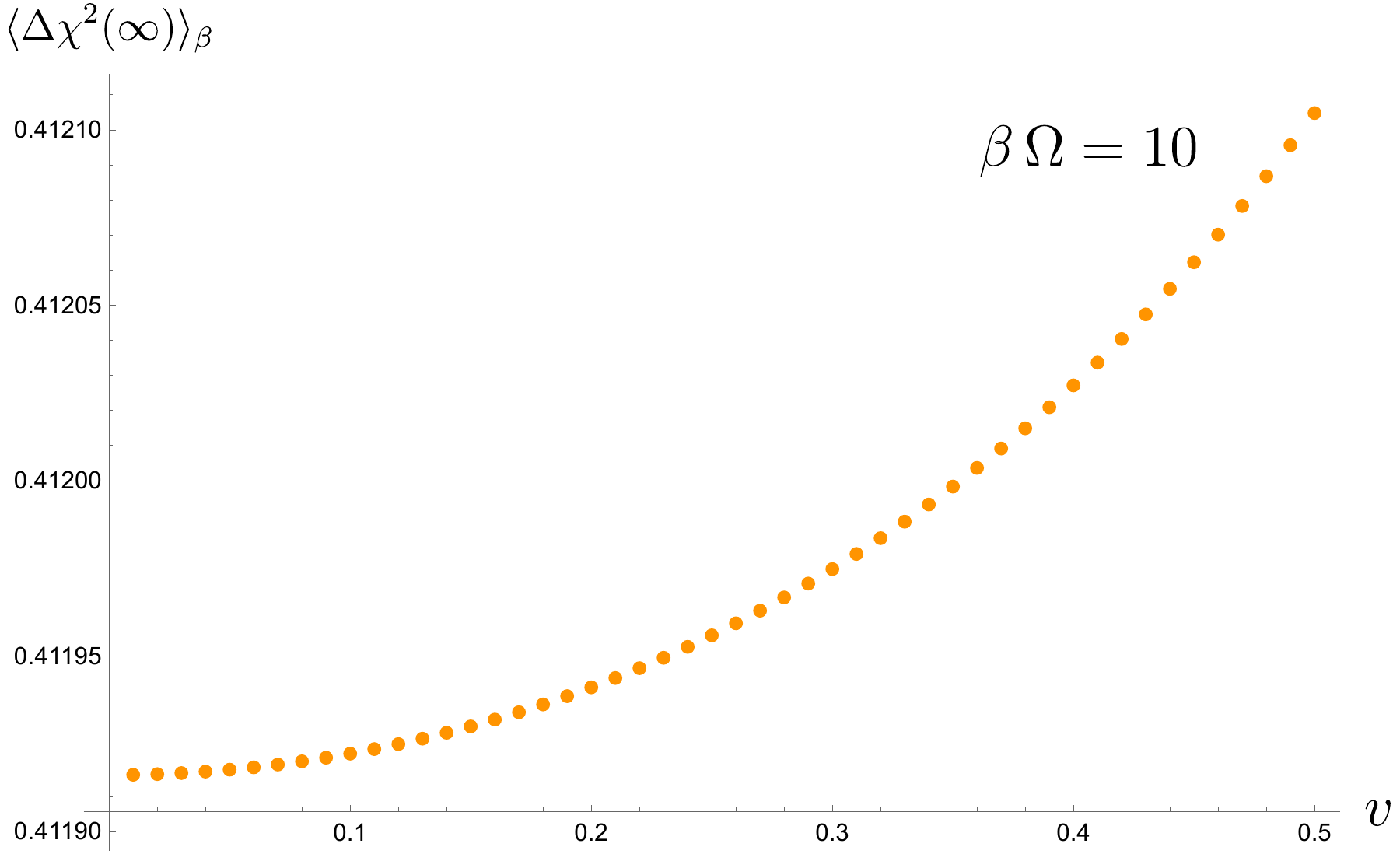}
        \caption{$\beta\Omega=10$}
        \label{Fi:graph2}
    \end{subfigure}
  
    \begin{subfigure}{0.45\linewidth}
        \centering
        \includegraphics[width=\linewidth]{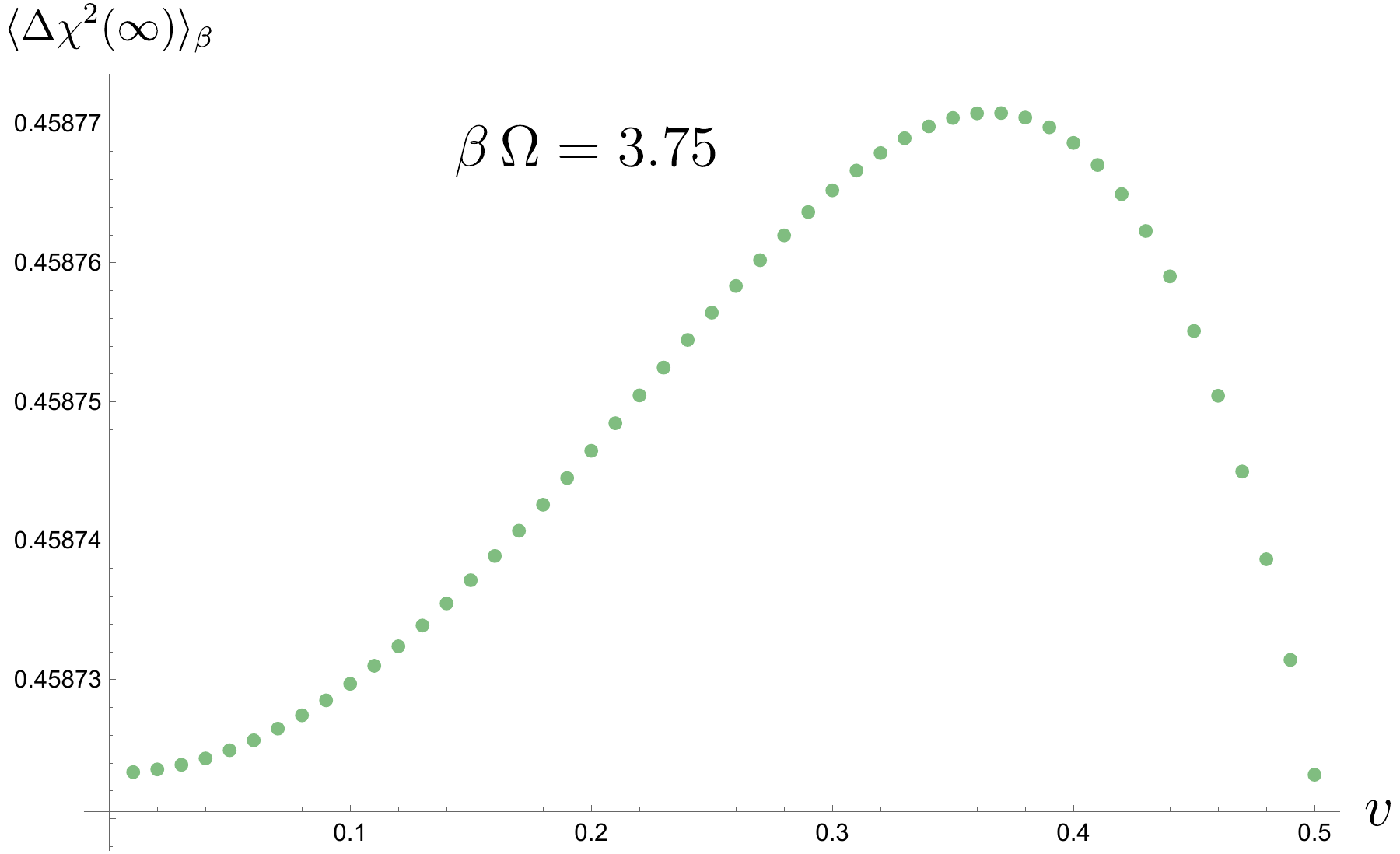}
        \caption{$\beta\Omega=3.75$}
        \label{Fi:graph3}
    \end{subfigure}
    \hfill
    \begin{subfigure}{0.45\linewidth}
        \centering
        \includegraphics[width=\linewidth]{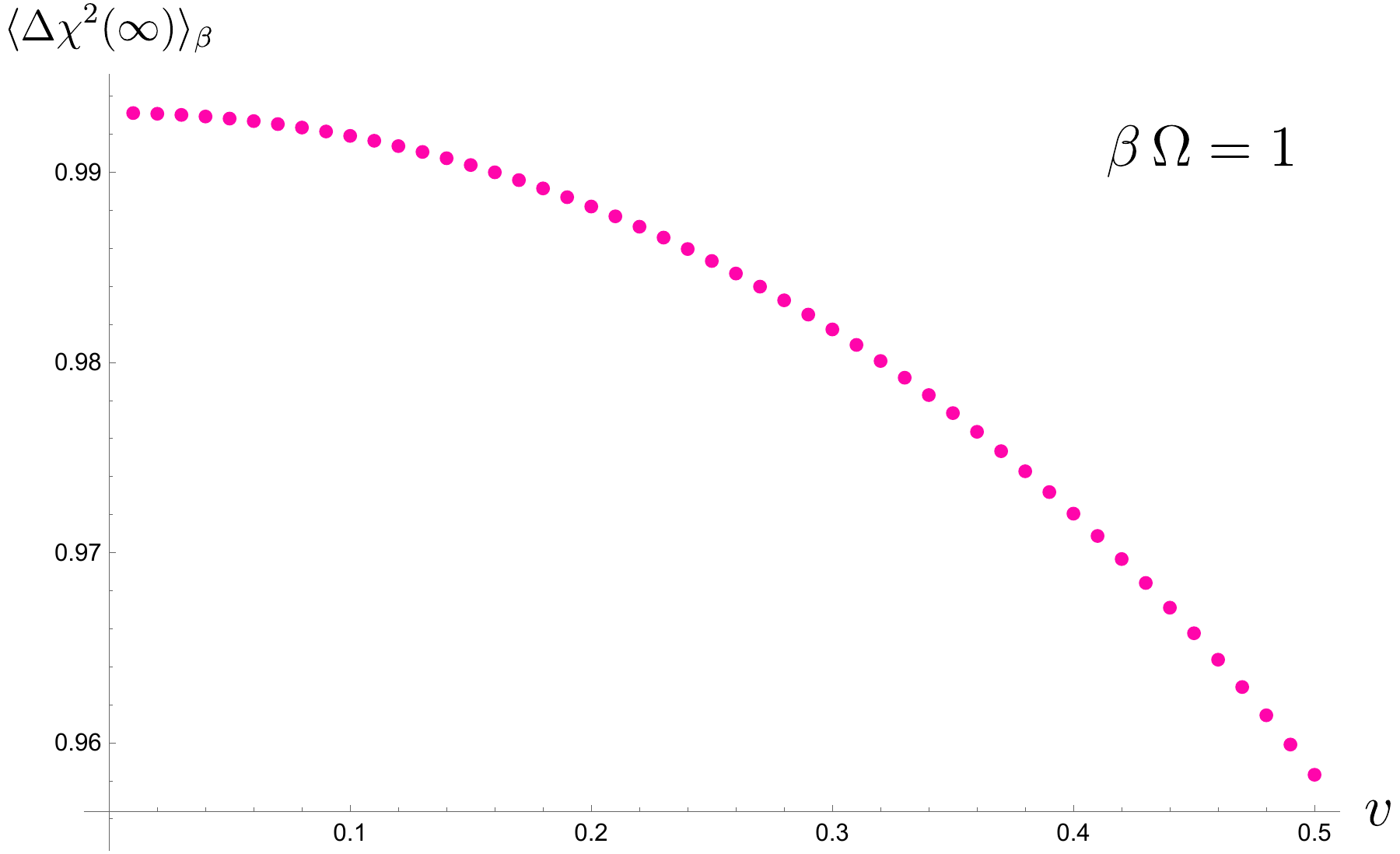}
        \caption{$\beta\Omega=1$}
        \label{Fi:graph4}
    \end{subfigure}

    \caption{The decoherence measure $\langle\Delta\chi^{2}\rangle$ plotted as a function of the atom velocity $v$ for different values of the bath temperature, characterized by $\beta\Omega$. At low temperatures (high $\beta\Omega$), the decoherence measure is relatively insensitive to changes in velocity, while at higher temperatures (low $\beta\Omega$), it exhibits a more pronounced dependence on velocity, decreasing as the atom moves faster. Here we have chosen $\Omega=1$, $m=1$, $\gamma=0.3$, and $\tau=15$ in the units of $\Omega$ or $\Omega^{-1}$. Note that the figures are not drawn to the same scale, so the variations may appear more significant than they actually are.}
    \label{Fi:decoherence_velocity}
\end{figure}

At low temperatures, $\widehat{G}_{\textsc{h}}^{(\phi)}(\varkappa)$ increases very mildly with the atom velocity, leading to a corresponding weak velocity dependence in the decoherence measure. To understand this behavior physically, we introduce the standard rapidity parameter $\eta$, characterizing the Lorentz boost, such that
\begin{align}
    &&v&=\tanh\eta\,,&\gamma_{v}&=\cosh\eta\,,&\gamma_{v}\,v&=\sinh\eta\,,\notag\\&\implies&\gamma_{v}(1+v)&=e^{\eta}=\sqrt{\frac{1+v}{1-v}}\,,&\gamma_{v}(1-v)&=e^{-\eta}=\sqrt{\frac{1-v}{1+v}}\,.
\end{align}
Using these relations, the comoving Hadamard function $\widehat{G}_{\textsc{h}}^{(\phi)}(\varkappa)$ in Eq.~\eqref{E:yiyjfg} can be re-expressed as
\begin{equation}
    \widehat{G}_{\textsc{h}}^{(\phi)}(\varkappa)=\frac{\lvert\varkappa\rvert}{4\pi}+\frac{1}{4\pi\beta\sinh\eta}\,\ln\frac{1-e^{\shortminus\beta\lvert\varkappa\rvert e^{+\eta}}}{1-e^{\shortminus\beta\lvert\varkappa\rvert e^{\shortminus\eta}}}\,.
\end{equation}
Excluding the ultra-relativistic regime where $e^{\pm\eta}\sim\mathcal{O}(1)$, we focus on the low-temperature or high-frequency limit, $\beta\lvert\varkappa\rvert=\lvert\varkappa\rvert/T\gg1$. In this regime, we can approximate the finite-temperature correction term via a leading-order expansion
\begin{align}\label{E:foertr}
    \widehat{G}_{\textsc{h},\beta}^{(\phi)}(\varkappa)=\frac{1}{4\pi\beta\sinh\eta}\,\Bigl(e^{\shortminus\beta\lvert\varkappa\rvert e^{\shortminus\eta}}-e^{\shortminus\beta\lvert\varkappa\rvert e^{+\eta}}\Bigr)\,,
\end{align}
Physically, this structure reflects Doppler-shifted thermal distributions. By defining a \textit{directional effective temperature} $T_{\text{eff}}=T\,e^{\eta}$, we see that for a given field mode frequency $\varkappa$ measured in the atom's rest frame, modes propagating against the atom's motion appear blue-shifted and hotter, whereas modes propagating in the direction of motion appear red-shifted and cooler.

These two competing contributions are captured by the respective exponential terms inside the parentheses in Eq.~\eqref{E:foertr}. Since the exponent of the first term contains the shrinking factor $e^{-\eta}$, making it less suppressed, it dominates over the second term, whose exponent contains the growing factor $e^{+\eta}$. Consequently, the net finite-temperature correction is positive and grows monotonically with rapidity $\eta$, or equivalently, with the atom's velocity $v$.

Intuitively, the blue-shifted incoming modes excite the internal degree of freedom more efficiently, while the red-shifted outgoing modes are less effective at de-exciting it. This asymmetry drives an overall increase in $\widehat{G}_{\textsc{h},\beta}^{(\phi)}(\varkappa)$, which in turn enhances the displacement variance $\langle\Delta\chi^{2}\rangle$. However, since this finite-temperature correction remains exponentially suppressed compared to the leading vacuum contribution $\widehat{G}_{\textsc{h},0}^{(\phi)}(\varkappa)$ in the regime $\beta \lvert\varkappa\rvert\gg1$, the velocity dependence is exceptionally weak, resulting in only a very mild enhancement of the decoherence measure.

At higher temperatures, the decoherence measure attains greater values and exhibits appreciable variations with the atom velocity, but it falls off with increasing velocity. In the high-temperature or low-frequency regime, $\beta\lvert\varkappa\rvert\ll1$, the finite-temperature correction is approximately
\begin{align}
    \widehat{G}_{\textsc{h},\beta}^{(\phi)}(\varkappa)\approx\frac{1}{4\pi\beta\sinh\eta}\,\ln\frac{\beta\lvert\varkappa\rvert e^{+\eta}+\frac{1}{2}\,\beta^{2}\lvert\varkappa\rvert^{2}\,e^{+2\eta}+\cdots}{\beta\lvert\varkappa\rvert e^{\shortminus\eta}+\frac{1}{2}\,\beta^{2}\lvert\varkappa\rvert^{2}\,e^{\shortminus2\eta}+\cdots}\approx\frac{\eta}{2\pi\beta\sinh\eta}\,,
\end{align}
if we keep only the leading order term in the expansion. In this case, the finite-temperature correction is approximately  proportional to $\dfrac{\eta}{\sinh\eta}$, and thus decreases with increasing atom velocity. Meanwhile, it is linearly proportional to the bath temperature $\beta^{-1}$, so the effect can be significantly amplified and typically dominates the vacuum contribution. Physically, in this regime, the internal degrees of freedom are well driven by the thermal fluctuations of the field, which effectively disrupt the quantum coherence in internal dynamics. Here, there is no longer an exponentially dominant hot forward contribution. However, the effect of relativistic aberration comes into play. It is also known as the \textit{headlight effect or relativistic beaming}, which causes the field modes to be concentrated in the forward direction of the atom motion. As the velocity grows, the frontward region gets increasingly hot and narrow, while other directions become colder than the bath rest-frame temperature. Thus, the atom integrates over a thermal landscape where most modes are heavily cooled, leading to a net reduction in $\widehat{G}_{\textsc{h},\beta}^{(\phi)}(\varkappa)$, and consequently a suppression of $\langle\Delta\chi^{2}\rangle$ with increasing atom velocity.

To formalize this geometric narrowing of the receiving cone, we examine the \textit{directional effective temperature},
\begin{align}
    T_{\text{eff}}(\vartheta)=\frac{T}{\gamma_{v}(1+v\,\varsigma)}\,,
\end{align}
which we have used in Eq.~\eqref{E:dbgrtz} with $\varsigma=\cos\vartheta$. The effective temperature is a function of the angle $\vartheta$ between the field mode and the atom velocity, as seen in the atom's rest frame. When $\varsigma=1$, the field mode is moving away from the atom, so the effective temperature reaches its minimum, $T_{\text{eff}}^{\text{min}}=T\,\sqrt{\dfrac{1-v}{1+v}}<T$. When the field mode is moving toward the atom, we have $\varsigma=-1$, and the effective temperature attains the maximum, $T_{\text{eff}}^{\text{max}}=T\,\sqrt{\dfrac{1+v}{1-v}}>T$. There exists a critical angle $\vartheta_{c}$ or a critical value of $\varsigma$ such that $T_{\text{eff}}(\vartheta_{c})=T$, which is given by
\begin{align}
    \varsigma_{c}=\cos\vartheta_{c}=-\frac{\gamma_{v}-1}{\gamma_{v}\,v}=-\tanh\frac{\eta}{2}\,.
\end{align}
Clearly $\theta_{c}>\pi/2$. As the rapidity $\eta$ increases, $\theta_{c}$ approaches $\pi$, demonstrating that the exceptionally hot incoming modes are tightly compressed into an increasingly narrow forward cone.

\section{Lab frame formulation of the internal dynamics, errors in ignoring relativistic effects }\label{S:eotge}

With fully relativistic treatments in the previous sections well displayed, in this section we point out problems, which exist but not so well noticed in the atomic, molecular and optical physics literature, where the relativistic effects on the internal degree of freedom are often ignored when the internal dynamics is formulated in the lab frame. In this case, the internal dynamics is governed by essentially the same equation of motion, except now the time coordinate is the lab time $t$ instead of the proper time $\tau$,
\begin{equation}
    \ddot{\chi}(t)+2\gamma\,\dot{\chi}(t)+\omega^{2}\,\chi(t)=\frac{\lambda}{m}\,\phi(\bm{z}(t),t)\,.
\end{equation}
Since the relativistic effects are ignored, we do not need to specify where the parameters of the internal degree of freedom are defined. On the other hand, we still keep the relativistic nature of the scalar field. The retarded Green's function and the Hadamard function of the field are now evaluated at the lab frame coordinates $(t,\bm{z}(t))$, where $\bm{z}(t)=\bm{v}\,t$ describes the uniform motion of the atom in the lab frame. The Hadamard Green's function is given by Eq.~\eqref{E:bgeit}. Thus, the results developed in the previous sections can be directly applied to this case, except that the time coordinate is now the lab time $t$ instead of the proper time $\tau$. The decoherence measure $\langle\Delta\chi^{2}(t)\rangle$ can be computed in a similar manner, and its dependence on the atom's velocity $v$ and the bath temperature $\beta$ can be analyzed accordingly.

Formally, the decoherence measure $\langle\Delta\chi^{2}(t)\rangle$ can be expressed as
\begin{align}\label{E:outuee}
    \langle\Delta\chi^{2}(t)\rangle&=d_{1}^{2}(t)\,\langle\chi^{2}(0)\rangle+d_{2}^{2}(t)\,\langle\dot{\chi}^{2}(0)\rangle+\frac{\lambda^{2}}{m^{2}}\int_{0}^{t}\!dt'\,dt''\;d_{2}(t')\,d_{2}(t'')\,G_{\textsc{h}}^{(\phi)}(t',\bm{z}(t');t'',\bm{z}(t''))\,,
\end{align}
where for example
\begin{equation}
    d_{2}(t)=\frac{1}{\Omega}\,e^{-\gamma t}\,\sin\Omega t\,,
\end{equation}
compared with Eq.~\eqref{E:bgrtir}. The intrinsic component of $\langle\Delta\chi^{2}(t)\rangle$, 
\begin{equation}
    \langle\Delta\chi^{2}(t)\rangle_{\text{intrinsic}}=d_{1}^{2}(t)\,\langle\chi^{2}(0)\rangle+d_{2}^{2}(t)\,\langle\dot{\chi}^{2}(0)\rangle\,,
\end{equation}
is independent of the environment, and depends only on the initial states of the internal degree of freedom. It essentially describes the intrinsic dynamics of the internal degree of freedom when the interaction with the environment is turned off, what we referred to earlier as quantum dispersion, an intrinsic quantum process. The coupling with the environment introduces an exponential decay of the \textit{intrinsic} component, characterized by the damping constant $\gamma$. The relaxation of the internal dynamics towards a steady state leads to a time-independent \textit{induced} component of $\langle\Delta\chi^{2}(t)\rangle$. Thus, at late times, the contribution from the intrinsic component becomes negligible, and the decoherence measure is dominated by the induced component.

The induced component of $\langle\Delta\chi^{2}(t)\rangle$,
\begin{align}
    \langle\Delta\chi^{2}(t)\rangle_{\text{induced}}&=\frac{\lambda^{2}}{m^{2}}\int_{0}^{t}\!dt'\,dt''\;d_{2}(t')\,d_{2}(t'')\,G_{\textsc{h}}^{(\phi)}(t',\bm{z}(t');t'',\bm{z}(t''))\notag\\ 
    &=\frac{\gamma}{\pi mv}\int_{0}^{\infty}\!d\kappa\;\coth\frac{\beta\kappa}{2}\int_{0}^{t}\!ds\int_{0}^{t}\!ds'\;\frac{\sin\kappa v(s-s')}{s-s'}\Bigl[e^{-i\kappa(s-s')}+e^{+i\kappa(s-s')}\Bigr]\,,
\end{align}
describes the influence of the environment on the internal degree of freedom, and is determined by the Hadamard function of the field evaluated along the atom trajectory. Thus, it depends on the properties of the environment, such as the bath temperature $\beta^{-1}$, the motion of the atom and the initial states, but it does not depend on the intrinsic properties of the internal degree of freedom. The induced component captures how the internal degree of freedom decoheres due to its interaction with the environment.

\begin{figure}
    \centering
    \begin{subfigure}{0.45\linewidth}
        \centering
        \includegraphics[width=\linewidth]{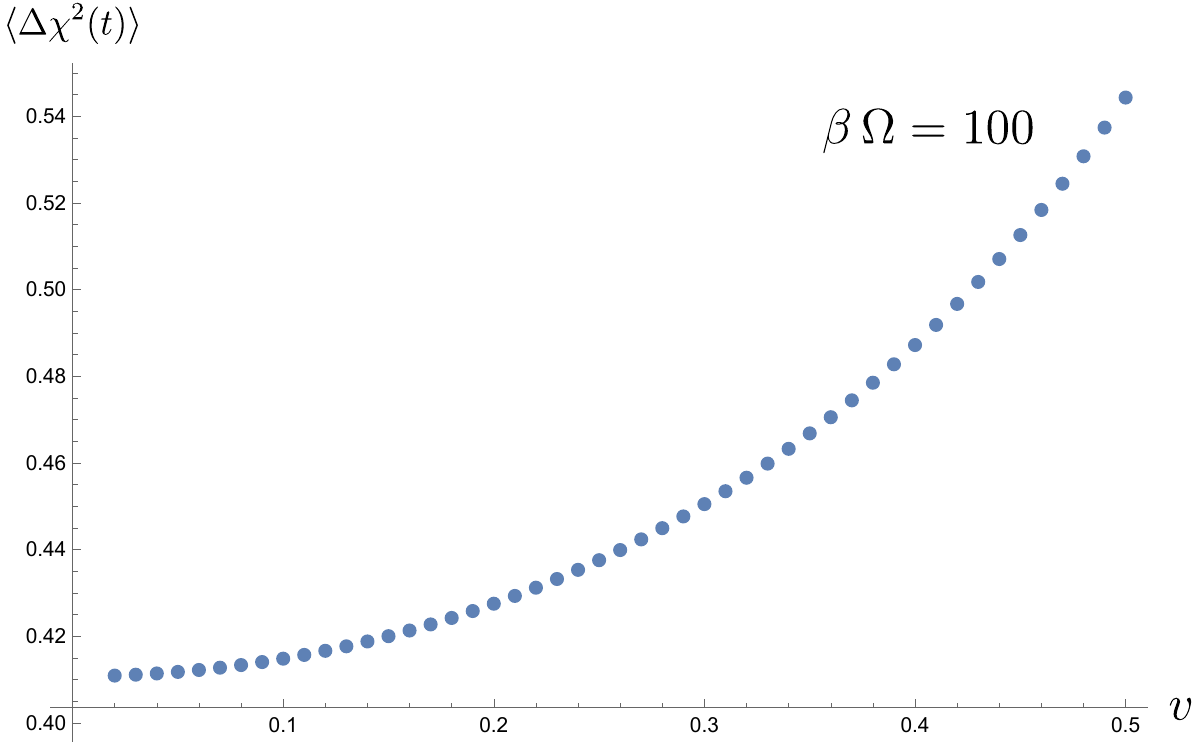}
        \caption{$\beta\Omega=100$}
        \label{Fi:graph5}
    \end{subfigure}
    \hfill
    \begin{subfigure}{0.45\linewidth}
        \centering
        \includegraphics[width=\linewidth]{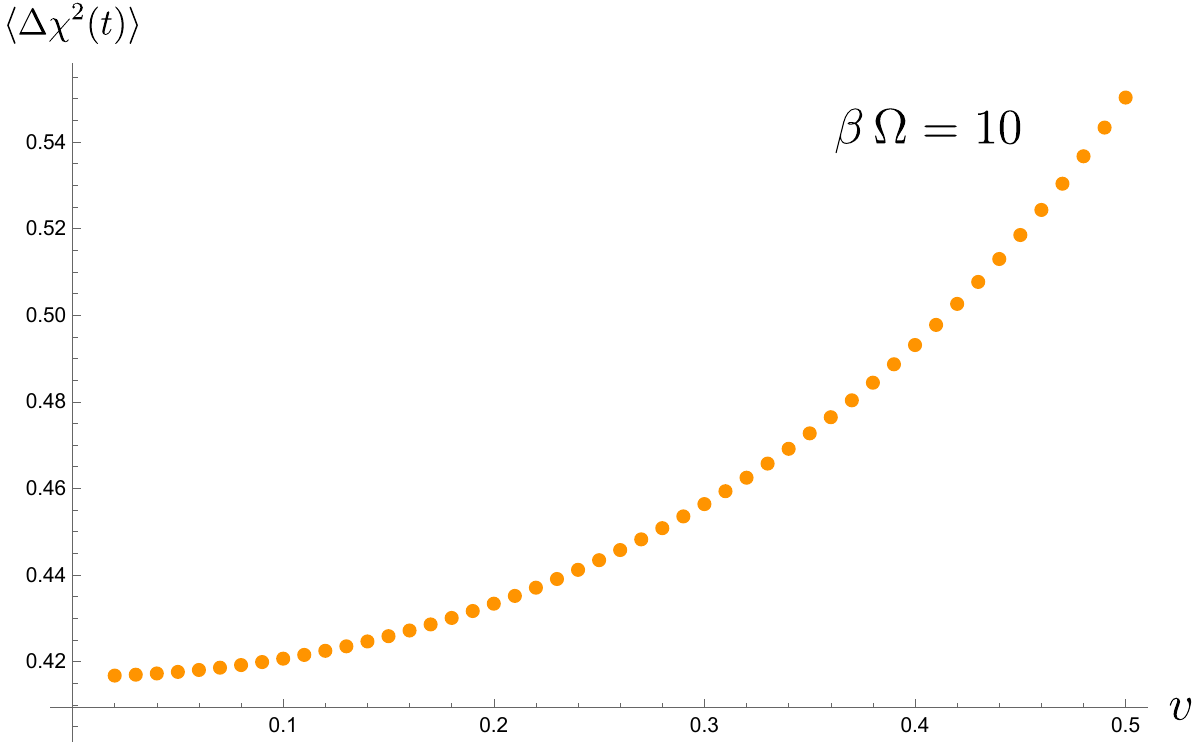}
        \caption{$\beta\Omega=10$}
        \label{Fi:graph6}
    \end{subfigure}
  
    \begin{subfigure}{0.45\linewidth}
        \centering
        \includegraphics[width=\linewidth]{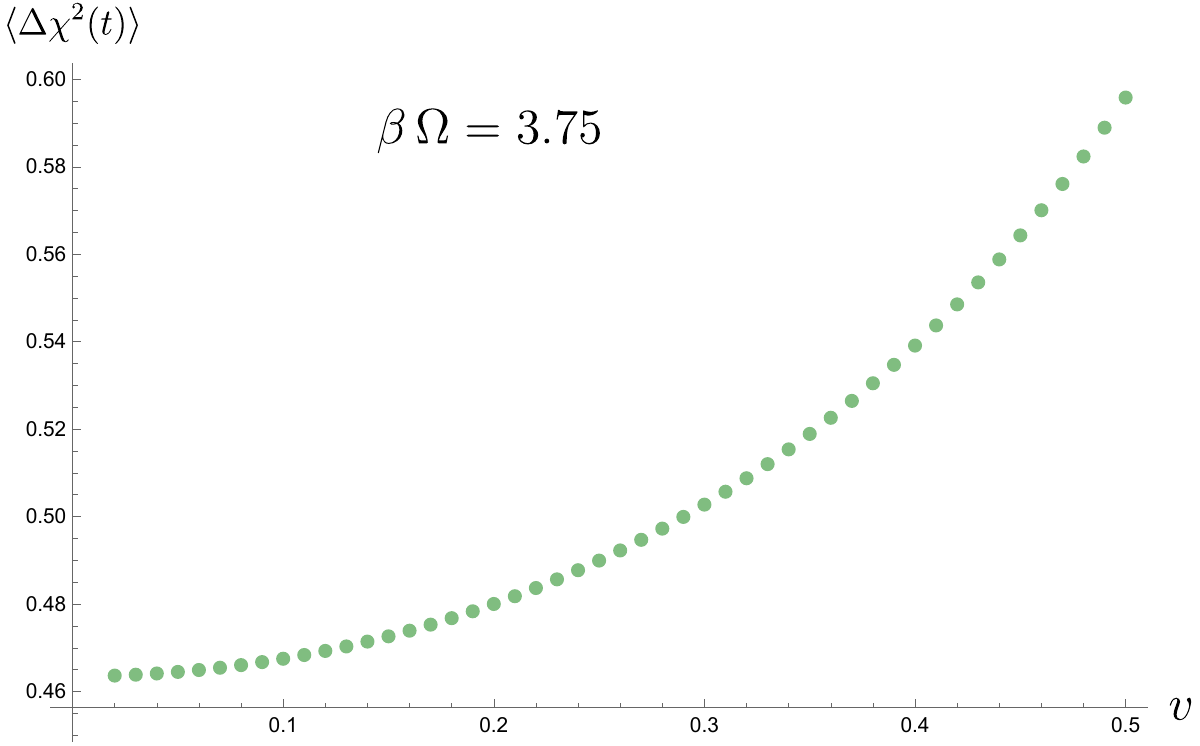}
        \caption{$\beta\Omega=3.75$}
        \label{Fi:graph7}
    \end{subfigure}
    \hfill
    \begin{subfigure}{0.45\linewidth}
        \centering
        \includegraphics[width=\linewidth]{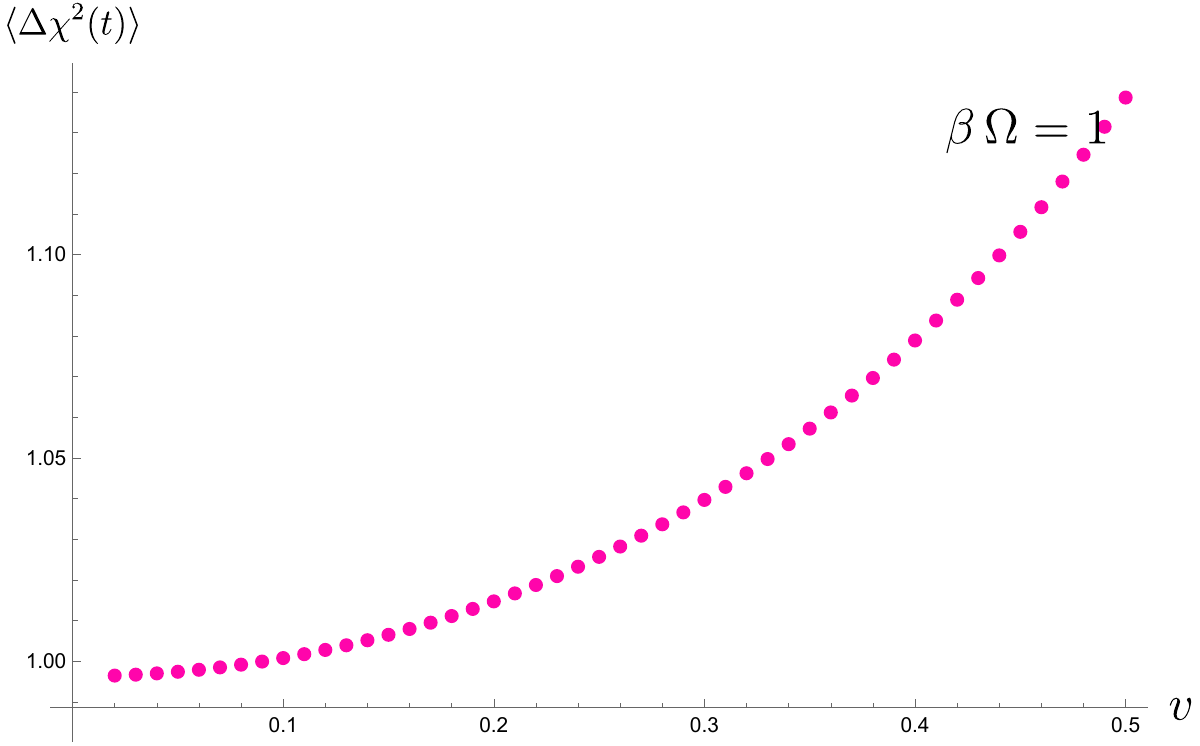}
        \caption{$\beta\Omega=1$}
        \label{Fi:graph8}
    \end{subfigure}
    \caption{Plotted here is the decoherence measure $\langle\Delta\chi^{2}\rangle$ as a function of the atom's velocity $v$ for different values of the bath temperature. In this case, relativity principle in the internal dynamics is incorrectly ignored. The same parameters are chosen as in Fig.~\ref{Fi:decoherence_velocity}.}
    \label{Fi:decoherence_velocity_lab}
\end{figure}
We are now in a position to compare the velocity dependence of the decoherence measure $\langle\Delta\chi^{2}(t)\rangle$ obtained  in this (improper) formulation versus that obtained from a proper treatment of the internal dynamics with full respect of relativity principles. As illustrated in Fig.~\ref{Fi:decoherence_velocity_lab}, the decoherence measure in the improper treatment increases monotonically with the atomic velocity $v$, a trend that becomes significantly more pronounced at elevated temperatures. This sharply contrasts with the correct proper-frame formulation, which predicted a suppression of decoherence with increasing velocity in the high-temperature regime. This discrepancy stems from an incomplete treatment of the internal degrees of freedom; notably, the dynamical variable $\chi$ in the current model is not a true Lorentz scalar. As expected, the results of both formulations converge in the ultra non-relativistic (non-moving) limit, $v\to0$ where $\gamma_v\to1$, as confirmed by the agreement between Fig.~\ref{Fi:decoherence_velocity} and Fig.~\ref{Fi:decoherence_velocity_lab}. But this is an extremely narrow range, say $\epsilon$, of the velocity span, essentially only when the atom is sitting still. Even by $2\epsilon$, the error has grown visibly significant. This shows the failure in the practice of adding relativistic ``corrections", at least in the present problem. But in our opinion, this prompts us to re-examine the viability of the accepted paradigm of introducing corrections to a non-relativistic calculation as a correct dictum.

We can see this from a comparison of the results shown in Fig.~\ref{Fi:graph5} which stand in stark contrast to those in Fig.~\ref{Fi:graph1} at non-vanishing velocities. In this case, $\beta\Omega=100$ effectively corresponds to the vacuum limit of the environment. Within this regime, Fig.~\ref{Fi:graph5} exhibits an unusually strong velocity-dependent decoherence, which stems from an artificial breakdown of Lorentz covariance.  The pathology in this model arises because the internal dynamics of the atom is not  formulated in a fully relativistic manner. Furthermore, while we naturally expect thermal fluctuations to severely disrupt quantum coherence in the high-temperature regime, the current model fails to account for relativistic aberration. Aberration anisotropically distorts the thermal bath perceived by a moving atom, an effect that becomes critical at relativistic velocities. Consequently, this model misrepresents the true behavior of the decoherence measure.  This comparison makes explicit the necessity of employing a strictly covariant treatment for the internal dynamics when evaluating decoherence in moving atoms.

\section{Fallacies in nonrelativistic or relativistic ``correction" treatments}\label{S:ebtort}

In this section, we examine the inconsistencies that arise in the non-relativistic calculations conventionally employed in atomic physics treatments. To elucidate these discrepancies and trace their physical origins, we compare the results of two distinct formulations: \textit{the fully relativistic treatment} (Model 1) expressed in Eq.~\eqref{E:ngldgj}, and \textit{the non-relativistic approximation} (Model 2) given by Eq.~\eqref{E:outuee}.
\[\resizebox{0.99\linewidth}{!}{$
\begin{aligned}
    &\text{Model 1:}  &\langle\Delta\chi^{2}(t)\rangle&=\mathsf{d}_{1}^{2}(\tau)\,\langle\chi^{2}(0)\rangle+\mathsf{d}_{2}^{2}(\tau)\,\langle\chi'^{2}(0)\rangle+\frac{\lambda^{2}}{m^{2}\gamma^{4}_{v}}\!\int_{0}^{t}\!dt'\,dt''\;\mathsf{d}_{2}(t')\,\mathsf{d}_{2}(t'')\,G_{\textsc{h}}^{(\phi)}(t'-t'')\,,\\
    &\text{Model 2:} &\langle\Delta\chi^{2}(t)\rangle&=d_{1}^{2}(t)\,\langle\chi^{2}(0)\rangle+d_{2}^{2}(t)\,\langle\dot{\chi}^{2}(0)\rangle+\frac{\lambda^{2}}{m^{2}}\!\int_{0}^{t}\!dt'\,dt''\;d_{2}(t')\,d_{2}(t'')\,G_{\textsc{h}}^{(\phi)}(t',\bm{z}(t');t'',\bm{z}(t''))\,,
\end{aligned}
$}\]
where $\mathsf{d}_{1}(t)=d_{1}(t/\gamma_{v})$ and $\mathsf{d}_{2}(t)=\gamma_{v}\,d_{2}(t/\gamma_{v})$, and examine how they converge in the limit $\gamma_v\to1$. At extremely low velocity, the dilation factor $\gamma_v$ is approximately given by $\gamma_v\sim 1+\dfrac{v^2}{2}$, so we have
\begin{align}
    \mathsf{d}_{1}(t)=d_{1}(t/\gamma_{v})&\sim d_{1}(t)-\frac{v^2\,t}{2}\,\dot{d}_1(t)+\mathcal{O}(v^3)\,,\\
    \mathsf{d}_{2}(t)=\gamma_{v}\,d_{2}(t/\gamma_{v})&\sim d_2(t)+\frac{v^2}{2}\bigl[d_2(t)-t\,\dot{d}_2(t)\bigr]+\mathcal{O}(v^3)\,.
\end{align}
The dilation factor falls off to unity very rapidly, and the intrinsic components in the above expressions converge quickly. The secular behavior in $t$ in the expansions does not pose a serious problem because both $d_1(t)$ and $d_2(t)$ decay exponentially fast with time. Thus, ignoring the dilation effect in the intrinsic components is justifiable in the limit $v\to0$.

However, more subtle contributions arise from the induced components in the individual integral expressions. In the second (incorrect) model, only the Doppler effect is included in the noise kernel, the measure of quantum field noise, which is of order $\mathcal{O}(v)$. While this captures the leading-order effect in the low-velocity limit and suffices when the field is initially in its vacuum state, it fundamentally fails to account for the relativistic effects of thermal fluctuations. Because a thermal state inherently defines a preferred reference frame, its physical effects are necessarily observer-dependent. As demonstrated by the first, fully covariant model, an observer moving relative to the thermal bath experiences not only a Doppler shift, but also a velocity-dependent effective temperature and a pronounced aberration effect. At higher velocities, these intertwined relativistic phenomena introduce intriguing, nontrivial features in the decoherence measure that the second model entirely misses.

Furthermore, when the motion becomes non-uniform, the time derivative of the Lorentz factor introduces terms proportional to $\bm{v}\cdot\bm{a}$, which are linear in $v$. The emergence of these non-inertial kinematic residues raises severe concerns regarding the validity of applying the second model to any form of accelerated motion.

To clearly illustrate these theoretical vulnerabilities, we summarize the comparison between the two models below. We deliberately evaluate a Lorentz-invariant measure of decoherence for the internal degree of freedom so that any violation of Lorentz covariance can be unambiguously discerned:
\begin{itemize}
    \item We compare two models describing the internal dynamics of an atom moving at constant speed: Model 1 based on the correct treatment in a fully covariant formulation, whereas Model 2 introduces ad-hoc kinematic modifications, often referred to as `relativistic corrections', to an atom at rest.
    \item Both models need to, as they do, agree in the ultra-nonrelativistic limit $v\to0$, essentially when the atom sits still.
    \item During the transient regime of the internal dynamics, Model 1 predicts a lower rate of relaxation and decoherence compared to Model 2. This is a direct consequence of relativistic time dilation.
    \item In the steady-state regime of the internal dynamics:
        \begin{itemize}
            \item If the field is initially prepared in a vacuum state, Model 1 yields a result that is expectantly independent of the atom's velocity, by virtue of Lorentz invariance. Starkly and erroneously different, the result predicted by Model 2 monotonically rises with increasing velocity.
            \item For a field initially in a thermal state, Model 2 predicts a monotonically increasing decoherence that rises rapidly with temperature, reflecting the disruption of internal coherence by thermal noise.
            \item Under the thermal condition, Lorentz invariance is broken, as the thermal bath defines a preferred rest frame.  Motional dependence of decoherence in Model 1 is not monotonic as foretold by Model 2. Instead, the relativistic aberration effect engages and plays a competing role, dynamically limiting the influence of the thermal fluctuations and thereby better preserving the coherence of the internal state of the moving atom.
        \end{itemize}
\end{itemize}

\subsection{Lesson to be learned here}

These comparative results clearly highlight the deep deficiencies in Model 2, serving as a stark warning against the prevalent methodology of refining a nonrelativistic baseline with piecemeal ``relativistic corrections." Investigating the dynamics of composite systems even at low relativistic speeds inherently demands the careful integration of relativistic kinematic effects. When modifications are introduced in an ad-hoc manner, based on specific, isolated needs rather than through a systematic, self-consistent framework based on relativistic covariance principles, fallacies are prone to occur.  As demonstrated in explicit calculations here, attempting to assemble the equations of motion with no regard paid to special relativity will unknowingly lead to artifacts and misleading conclusions, such as the spurious velocity dependence in the vacuum state. To rigorously insulate the theory against such pitfalls, the commonly adopted theoretical paradigm of first obtaining a nonrelativistic description and then adding nonrelativistic corrections must be inverted: the dynamics must first be derived and regularized within a fully covariant relativistic framework (Model 1). Only after the exact, internally consistent equations of motion are established should low-velocity limits be extracted. Only in this way can all the relativistic effects, time dilation, aberration, non-inertial mass shifts, be systematically preserved.

\subsection{Fallacies of similar treatments in related areas}

Similar issues exist in other areas. We mention just one closely related to what is addressed here. In some phenomenological models of quantum friction, a tempting shortcut is to incorporate relativistic motion into a nonrelativistic interaction solely through the Doppler substitution $\omega\mapsto\omega-\bm{v}\cdot\bm{k}$. Although the Doppler shift itself is physically well motivated, applying it selectively to rest-frame expressions does not generally provide a complete relativistic treatment of the field's Green's functions and quantum state, particularly for a thermal environment~\cite{PapAna20}. Such partial modifications, when applied inconsistently to the field correlations and response functions, can introduce artificial discrepancies between the noise and dissipation kernels, potentially leading to spurious conclusions about fluctuation-dissipation relations~\cite{JohHu02}. Indeed, studies of quantum friction have demonstrated how inconsistent approximations to fluctuations, dissipation, and back-reaction can produce predictions incompatible with steady-state energy balance~\cite{ReiEtAl20}. These examples underscore the importance of deriving relativistic corrections self-consistently from the underlying field theory, rather than introducing them selectively into an otherwise nonrelativistic framework.

\vskip18pt
\noindent{\bf Acknowledgments} 
J.-T. Hsiang is supported by the National Science and Technology Council of Taiwan, R.O.C. under Grant Nos.~113-2112-M-011-001-MY3 and 115-2112-M-011-002-.

\newpage
\bibliography{MotDec}

\end{document}